\documentclass[]{jfm}

\usepackage{graphicx}
\usepackage{newtxtext}
\usepackage{newtxmath}
\usepackage{natbib}
\usepackage{hyperref}
\usepackage{cancel}
\usepackage{comment}
\usepackage{siunitx}
\hypersetup{
   colorlinks = true,
   urlcolor   = blue,
   citecolor  = blue,
   allcolors  = blue
}

\newcommand{\RomanNumeralCaps}[1]
\linenumbers



\shorttitle{Buoyancy correction of modal instabilities in stratified boundary layers}
\shortauthor{P.~C.~Boldini et al.}

\title{First-order buoyancy correction of modal instabilities in stratified boundary layers}

\author{P.~C.~Boldini\aff{1}$\dagger$,
 R.~Hirai\aff{1}, B.~Bugeat\aff{2}, R.~Pecnik\aff{1} \corresp{\email{p.c.boldini@tudelft.nl}, r.pecnik@tudelft.nl}}

\affiliation{\aff{1}Process and Energy Department, Delft University of Technology, Leeghwaterstraat 39, 2628 CB Delft, the Netherlands
            \aff{2}School of Engineering, University of Leicester, University Road, Leicester, LE1 7RH, United Kingdom}

\begin{document}
\maketitle

\begin{abstract}
We present a perturbation-based framework that captures buoyancy effects on modal instabilities in stratified boundary-layer flows within the fully compressible, non-Oberbeck--Boussinesq formulation. Treating the Richardson number as a small parameter and recasting the stability problem into an adjoint-residual form, we derive a first-order correction for the eigenvalues using only the neutrally buoyant eigenvalue problem. This eliminates the need to re-solve the eigenvalue problem at each stratification level. For ideal-gas boundary layers, the framework accurately predicts how stable and unstable stratification modifies Tollmien--Schlichting waves, from growth rates and eigenfunctions to $N$-factors, holding across a wide range of Prandtl numbers, temperature ratios, and Mach numbers. Notably, the buoyancy sensitivity varies strongly with Prandtl number, revealing that for a given Richardson number, buoyancy can switch from destabilising to stabilising depending on the fluid. Beyond ideal-gas conditions, we apply the first-order buoyancy correction to strongly stratified boundary layers with supercritical fluids, where the phase relationship between density and velocity perturbations determines whether buoyancy stabilises or destabilises the underlying instability. The resulting $N$-factors demonstrate, for the first time, that buoyancy significantly affects transition predictions under pseudo-boiling conditions.
\end{abstract}

\begin{keywords}
\end{keywords}


\section{Introduction}
\label{sec1:intro}
Wall-bounded stratified flows are central to both geophysical and engineering problems. They occur, for example, in atmospheric and oceanic boundary layers, as well as in engineering systems such as heat and mass transfer equipment. In these flows, wall-normal temperature gradients cause density variations and, consequently, buoyancy forces, which can affect the shear-driven boundary-layer dynamics. Extensive work has been carried out on wall-bounded stratified turbulence, as summarised in \citet{Zonta2018,Caulfield2021}. In contrast, the hydrodynamic instability of stratified boundary layers remains largely unexplored. Instability in these flows may originate from Tollmien--Schlichting (TS) waves or from buoyancy-driven convection. Depending on the level of stratification, these mechanisms can compete, interact, or merge into mixed modes \citep{Gage1968,Sameen2007}. When buoyancy forces overcome viscous and thermal diffusion, convection-like instabilities may develop \citep{Hall1992,Carriere1999,Hirata2015}, while in the shear-dominated regime, TS waves remain the primary driver of transition \citep{Gebhart1973}. We confine our attention to the shear-dominated regime rather than to mixed or purely buoyancy-driven convection.

Classical analyses of stratified shear flows often rely on the inviscid Taylor--Goldstein equation, derived under the Oberbeck--Boussinesq (OB) approximation \citep{Drazin2004}. In this inviscid framework, stability is assessed through the local gradient Richardson number $Ri_\mathit{g}=N^2/S^2$, which compares the stabilising effect of buoyancy, expressed by the Brunt--Väisälä frequency $N$, with the destabilising effect of the wall-normal shear $S$. The Miles--Howard theorem states that a parallel, steady, inviscid, and stably stratified shear flow remains asymptotically stable if $Ri_\mathit{g}(y)>1/4$ holds everywhere, where $y$ is the wall-normal coordinate \citep{Miles1961,Howard1961}. 

Kelvin--Helmholtz-type mechanisms are present in free-shear flows and geophysical contexts, but generally not in wall-bounded boundary layers, where TS waves usually dominate \citep{Chen2016}. Note that an exception was reported by \citet{Boldini2025b}, who observed billow-like structures, resembling Kelvin--Helmholtz instability, in the near-wall region of boundary layers with fluids at supercritical pressure. Stable stratification damps TS waves and increases the critical Reynolds number and corresponding frequency, whereas sufficiently unstable stratification can promote convection-like disturbances that compete with or even overtake TS waves \citep{Wu1976,Sameen2007}. In this scenario, mixed or buoyancy-dominated modes are present. The viscous counterpart of the Taylor--Goldstein equation is the Orr--Sommerfeld equation with buoyancy coupling: stratification modifies, but does not generally suppress, viscous instabilities \citep{Drazin2004}. No general viscous criterion exists, with the exception of the asymptotic-suction boundary layer at $Pr=1$, where a sufficient condition for stability is $Ri_\mathit{g}(y)>0.0554$ \citep{Gage1971}. 

Extracting general trends remains challenging: experimental and numerical studies demonstrate that the linear stability characteristics, such as neutral curves and critical Reynolds numbers, vary systematically with Richardson (or Froude) and Prandtl numbers \citep{Biau2004,Facchini2018,Gal2021,Variale2024}. Yet, a unifying theoretical framework to predict these trends has not emerged so far. Recent studies have focused on the effect of stratification on modal stability in horizontal boundary layers. \citet{Parente2020} carried out stability analyses across a range of Prandtl, Reynolds, and Richardson numbers, revealing stabilisation by stable stratification, while the effect of Prandtl number is non-monotonic: stabilising up to $Pr\approx 7$ (i.e.~water) before destabilising at higher $Pr$. Stratification also reshapes the disturbance eigenfunction: \citet{Thummar2024} reported an upward shift of its peak with increasing $Ri$. Complementing these results, \citet{Hamada2023} used resolvent analysis to show that buoyancy can transfer energy from thermal to kinetic perturbations and that the strongest amplification occurs in continuous modes, making the boundary layer more receptive to free-stream forcing than to TS modes. 

Although the influence of buoyancy on modal instabilities, e.g.~TS waves, has been increasingly investigated in stratified boundary layers, a systematic quantitative characterisation is still lacking. Here we introduce an efficient perturbative framework that captures buoyancy effects on hydrodynamic instabilities without repeatedly solving the eigenvalue problem of the buoyant boundary layer at each stratification level. First-order corrections to viscous eigenvalues and eigenvectors are derived for both stable and unstable stratification by recasting the problem in residual form and treating the Richardson number as a perturbation parameter. Unlike previous studies under the OB approximation, we develop the perturbative framework under the fully compressible, non-Oberbeck--Boussinesq (NOB) formulation to capture stratification effects on modal instability without restricting to small density variations. \citet{Ke2024} recently demonstrated that NOB effects significantly alter boundary-layer stability characteristics as temperature difference increases; however, their study was limited to ideal-gas conditions. In this work, we apply the perturbative framework to non-ideal-gas conditions \citep{Guardone2024}, where the interplay between stratification and modal instabilities has not been previously investigated. This is particularly relevant at near-critical conditions above the vapour-liquid critical point \citep{Li2025}, where minimal temperature variations induce sharp changes in density and transport properties. Such conditions enable exceptional heat-transfer performance exploited in advanced energy systems \citep{Brunner2010,Liu2019}.

The paper is organised as follows. \S\,\ref{sec:2} introduces the governing equations for the linear stability analysis of buoyant boundary layers. The first-order buoyancy correction framework is derived in \S\,\ref{sec:3}. Its accuracy is assessed in \S\,\ref{sec:4} for both stable and unstable stratification, including parametric variations of the base flow. \S\,\ref{sec:5} extends the perturbation framework to boundary-layer flows of carbon dioxide ($\mathrm{CO}_2$) at supercritical pressure. Finally, conclusions are drawn in \S\,\ref{sec:6}.

\section{Methodology}
\label{sec:2}
\subsection{Governing equations}
\label{sec:2a}

The fully compressible Navier--Stokes (NS) equations are written in conservative and non-dimensional form following \citet{Boldini2025} as
\begin{subequations}\label{eq:NS}
\begin{gather}
\frac{\partial \rho}{\partial t} + \frac{\partial (\rho u_\mathit{j})}{\partial x_\mathit{j}} = S_\rho, \nonumber \\
\frac{\partial (\rho u_\mathit{i})}{\partial t} + \frac{\partial (\rho u_\mathit{i} u_\mathit{j} + p \delta_\mathit{ij} - \tau_\mathit{ij})}{\partial x_\mathit{j}} = S_\mathit{i}, \label{eq:NS:mom} \tag{\theequation$a$--$c$} \\
\frac{\partial (\rho e_0)}{\partial t} + \frac{\partial \big[(\rho e_0 + p)u_j + q_\mathit{j} - u_\mathit{i} \tau_\mathit{ij}\big]}{\partial x_\mathit{j}} = S_\mathit{e}, \nonumber
\end{gather}
\end{subequations}
where $x_\mathit{j}=x^*_\mathit{j}/\delta^* = \left(x,y,z\right)$ are the Cartesian coordinates in the streamwise, wall-normal, and spanwise directions, respectively, $u_\mathit{j} = u^*_\mathit{j}/u^*_{\infty} = \left(u,v,w\right)$ are the corresponding velocity components, $\rho=\rho^*/\rho^*_{\infty}$ is the density, $p=p^*/(\rho^*_{\infty} {u^{*2}_{\infty}})$ is the pressure, and $e_0=e^*_0/u^{*2}_{\infty}=e+u_\mathit{j} u_\mathit{j}/2$ is the specific total energy, with $e=e^*/u^{*2}_{\infty}$ as the specific internal energy. The viscous stress tensor $\tau_\mathit{ij}$ is calculated as $\tau_\mathit{ij}=\lambda/Re \, \partial u_\mathit{k}/\partial x_\mathit{k} \delta_\mathit{ij} + \mu/Re \, (\partial u_\mathit{i}/\partial x_\mathit{j} + \partial u_\mathit{j}/\partial x_\mathit{i})$, where $\mu=\mu^*/\mu^*_{\infty}$ is the dynamic viscosity, $\lambda=-2/3 \mu $ is Lam\'e's constant with zero bulk viscosity (Stokes' hypothesis), and $\delta_\mathit{ij}$ is the Kronecker delta. The heat flux vector $q_\mathit{j}$ follows Fourier's law as $q_\mathit{j}=-\kappa/(Re Pr_\infty Ec_\infty) \, \partial T/\partial x_\mathit{j}$, where $\kappa=\kappa^*/\kappa^*_{\infty}$ is the thermal conductivity and $T=T^*/T^*_{\infty}$ is the fluid temperature. Dimensional quantities are denoted by $(\cdot)^*$, and $(\cdot)^*_\infty$ refers to free-stream conditions. The corresponding characteristic parameters are
\begin{subequations}
\begin{gather}
Re=\frac{\rho^*_{\infty}u^*_{\infty}\delta^*}{\mu^*_{\infty}}, \quad 	Ec_\mathit{\infty}=\frac{{u^{*2}_\infty}}{c^*_\mathit{p,\infty}T^*_\infty}, \quad 
Pr_\mathit{\infty}=\frac{c^*_{\mathit{p},\infty} \mu^*_{\infty}}{\kappa^*_\infty}, \tag{\theequation$a$--$c$}
\label{eq:nondimnumbers}
\end{gather}
\end{subequations}
where $c^*_{\mathit{p},\infty}$ is the specific isobaric heat capacity, $Re$ is the Reynolds number based on the local Blasius length scale $\delta^*=(\mu^*_\infty x^*/(\rho^*_\infty u^*_\infty))^{1/2}$, $Ec_\infty$ is the Eckert number, and $Pr_\infty$ is the Prandtl number. The Mach number $M_\infty=u^*_\infty/a^*_\infty$, with $a^*_\infty$ the speed of sound, is obtained from $Ec_\infty$.

Buoyancy forces enter the NS equations \eqref{eq:NS} as the source-term vector $\mathbf{S}=(S_\rho,S_x,S_y,S_z,S_e)$, defined as
\begin{equation}
    \mathbf{S}=\left(0,0,e_y\dfrac{\rho}{Fr^2},0,e_y\dfrac{\rho v}{Fr^2}\right)^T, \quad \text{with} \, Fr=\dfrac{u^*_{\infty}}{\sqrt{ g^* \delta^*}},
    \label{eq:source}
\end{equation}
where $Fr$ denotes the Froude number. The gravitational acceleration $g^*$ acts along the wall-normal $y$-direction, and $e_y \in \{-1, 1\}$ denotes the directional sign of gravity. In the present NOB formulation (see also \citet{Zonta2018}), buoyancy effects arise directly from the large density variations in the boundary layer, which are fully retained in both inertia and buoyancy without linearisation about a reference state. A convenient measure of NOB effects in this context is the 
relative density variation $\Delta\rho^*/\rho^*_\infty$, where $\Delta \rho^*=\rho^*_\mathit{w}-\rho^*_\infty$ denotes the density difference between the wall and the free stream, with subscript $(\cdot)_\mathit{w}$ denoting wall quantities. It directly quantifies the strength of stratification and, for small density variations, may be related to the Atwood number, $At \approx \Delta\rho^*/(2\rho^*_\infty)$, commonly used in stratified shear-flow studies \citep{Guha2018}. In combination with the Froude number, we therefore parametrise buoyancy effects by the Richardson number defined as:
\begin{equation}
  Ri=-e_y\frac{\Delta\rho^*g^* \delta^*}{\rho^*_\infty u^{*2}_\infty}.
  \label{eq:Ri}
\end{equation}
For weakly or moderately stratified boundary-layer flows, 
$|Ri| \ll 1$ corresponds to 
$\big|\Delta\rho^*/(\rho^*_\infty Fr^2)\big| \ll 1$ in \eqref{eq:Ri}. When additionally $|\Delta\rho^*/\rho^*_\infty| \ll 1$, the classical OB approximation \citep{Parente2020,Hamada2023} is recovered, in which the density and temperature difference $\Delta T^*=T^*_\mathit{w}-T^*_\infty$ are related by $\Delta \rho^*/\rho^*_\infty \approx -\beta^*_\infty \Delta T^*$, where $\beta^*_\infty$ is the free-stream thermal expansion coefficient.

With the definition of the Richardson number \eqref{eq:Ri}, the source-term vector \eqref{eq:source} can be reformulated. The hydrostatic background pressure $p^*_\mathit{h}$ across the boundary layer scales as $\Delta p^*_\mathit{h} \sim \Delta\rho^* g^* \delta^*=Ri \,\rho^*_\infty u^{*2}_\infty$, i.e. its variation is of order $\mathcal{O}(Ri)$ in non-dimensional form. For weakly and moderately stratified boundary-layer flows with $|Ri|\ll 1$, this variation is small compared with the inertial pressure scale $\rho^*_\infty u^{*2}_\infty$ and we therefore subtract $p^*_\mathit{h}$ from the total pressure (see also \citet{Boldini2025}). The source-term vector then becomes
\begin{equation}
    \mathbf{S}=\left(0,0,-Ri\dfrac{\rho^*_\infty}{\Delta\rho^*}(\rho-1),0,-Ri\dfrac{\rho^*_\infty}{\Delta\rho^*}(\rho-1)v\right)^T,
    \label{eq:source2}
\end{equation}
which explicitly retains the local density deviation from the free-stream conditions through the factor $(\rho-1)$. According to the definition of the Richardson number \eqref{eq:Ri}, $Ri>0$ and $Ri<0$ correspond to stable and unstable stratification, respectively, regardless of the direction of gravity. 

Both ideal and non-ideal gas conditions are considered in this study. Under the ideal-gas assumption, the pressure satisfies the ideal-gas law $p=\rho R_\mathit{g} T$, where $R_\mathit{g}$ is the specific gas constant, and the transport coefficients $\mu$ and $\kappa$ are evaluated using Sutherland’s law. For non-ideal gas effects \citep{Guardone2024}, the NIST REFPROP library \citep{Lemmon2013} is used to obtain the multiparameter equation of state $p=p(\rho,T)$, the caloric relation $e = e(\rho,T)$, and the transport coefficients.

\subsection{Linearised stability equations for stratified boundary layers}
\label{sec:2b}
Two-dimensional waves are considered under the locally parallel flow assumption \citep{Mack1984}. The flow field $\mathbf{Q}=(\rho,u,v,T)^T$ is decomposed into a base flow $\mathbf{\bar{Q}}$ and a perturbation $\mathbf{q}$, and is substituted into the NS equations \eqref{eq:NS}. After subtracting the base flow and neglecting nonlinear terms, the linearised stability equations read
\begin{flalign}\label{eq:LinearNS}
  \begin{aligned}
    &\mathcal L_{t} \frac{\partial \mathbf{q}}{\partial t}  +
    \mathcal L_{x}  \frac{\partial \mathbf{q}}{\partial x}+
    \mathcal L_{y}  \frac{\partial \mathbf{q}}{\partial y} +
    \mathcal L_{q} \mathbf{q} + \mathcal{L}_{q,\mathbf{S}} \, \mathbf{q}\\
    &+\mathcal V_{xx} \frac{\partial^2 \mathbf{q}}{\partial x^2} +
    \mathcal V_{xy} \frac{\partial^2 \mathbf{q}}{\partial x \partial y} +
    \mathcal V_{yy} \frac{\partial^2 \mathbf{q}}{\partial y^2} 
    =0,
   \end{aligned}
\end{flalign}
where $\mathcal{L}_\mathit{t}$, $\mathcal{L}_\mathit{x}$, $\mathcal{L}_\mathit{y}$, $\mathcal{L}_\mathit{q}$, $\mathcal{L}_\mathit{q,\mathbf{S}}$ (containing the linearised contributions of $\mathbf{S}$ in \eqref{eq:source2}), $\mathcal{V}_\mathit{xx}$, $\mathcal{V}_\mathit{xy}$, and $\mathcal{V}_\mathit{yy}$ depend on the base flow and are given in Appendix~\ref{sec:appA}. The perturbation is expressed in normal-mode form as
\begin{equation}\label{eq:Fourier}
\mathbf{q}(x,y,t)=\hat{\bold{q}}(y)\exp[\mathrm{i}(\alpha x-\omega t)] + \text{c.c.},
\end{equation}
where $\hat{\bold{q}}(y)$ is the perturbation eigenfunction, $\alpha$ is the streamwise wavenumber, $\omega$ is the angular frequency, and c.c.~denotes the complex conjugate. In the spatial framework, $\alpha$ is complex, and its imaginary part $\Im\{\alpha\}$ represents the spatial growth rate, with modal amplification for $-\Im\{\alpha\}>0$. Substituting \eqref{eq:Fourier} into \eqref{eq:LinearNS} yields
\begin{equation}\label{eq:Eig_problem3}
\mathcal{A} \hat{\bold{q}} = \alpha \mathcal{B} \hat{\bold{q}} + \alpha^2\mathcal{V}_\mathit{xx} \hat{\bold{q}},
\end{equation}
where 
\begin{subequations}
\begin{gather}\label{eq:Eig_problem4}
\mathcal{A} = -\mathrm{i} \omega \mathcal{L}_\mathit{t} + \mathcal{L}_\mathit{y}D + \mathcal{L}_\mathit{q}+\mathcal{L}_\mathit{q,\mathbf{S}}+ D^2 \mathcal{V}_\mathit{yy}, \quad \mathcal{B} = - \mathrm{i} \mathcal{V}_\mathit{xy}D - \mathrm{i} \mathcal{L}_\mathit{x} , \tag{\theequation$a$,$b$}
\end{gather}
\end{subequations}
and $D=\mathrm{d}/\mathrm{d}y$ denotes the wall-normal derivative operator. Equation \eqref{eq:Eig_problem3} constitutes a quadratic eigenvalue problem that depends on $\mathbf{\bar{Q}}$, obtained from the self-similar boundary-layer equations for ideal-gas \citep{Schlichting2003} and non-ideal-gas \citep{Ren2019b} conditions. We follow \citet{Parente2020,Hamada2023} in assuming that the buoyancy force does not affect the (laminar) base flow $\mathbf{\bar{Q}}$. To solve the eigenvalue problem \eqref{eq:Eig_problem3}, recast as a linear eigenvalue problem (see \citet{Boldini2024}), the system is discretised using $N_\mathit{y}$ Chebyshev collocation points and Chebyshev differentiation matrices (the $D$ operator) \citep{Malik1990}.
At the wall ($y=0$), the fluctuations in streamwise and wall-normal velocities, $u$ and $v$, are set to zero in accordance with the no-slip condition, and the temperature fluctuation $T$ is set to zero, consistent with an isothermal wall. In the free stream ($y\to\infty$), Dirichlet boundary conditions are imposed for $u$, $v$, and $T$.

\section{Derivation of perturbation theory for stratified boundary layers}
\label{sec:3}
We develop a first-order perturbation framework following \citet{Kato1995} to assess and predict the stability of buoyant boundary layers, treating the Richardson number as a small parameter. To this end, we rewrite the eigenvalue problem \eqref{eq:Eig_problem3} in a residual-operator form suitable for perturbation analysis. Since the matrix $\mathcal{B}$ in \eqref{eq:Eig_problem4} depends explicitly on $\alpha$, we introduce
\begin{equation}
  \mathcal{R}=\mathcal{R}(\mathbf{\bar{Q}},\alpha,\omega,Re,Ri)=\mathcal{A}-\alpha \mathcal{B}-\alpha^2\mathcal{V}_\mathit{xx},
  \label{eq:25}
\end{equation}
where $\mathcal{R}$ is the residual operator. Our aim is to derive an analytic expression for the first-order buoyancy correction to the eigenvalue $\alpha$. For fixed $\mathbf{\bar{Q}}$, $Re$, and $\omega$, we hence write $\mathcal{R}=\mathcal{R}(\alpha,Ri)$, so that the eigenvalue problem \eqref{eq:Eig_problem3} of the buoyant boundary layer can be written as:
\begin{equation} \label{eq:residual_wbuo}
    \mathcal{R}(\alpha,Ri)\,\hat{\bold{q}}=0.
\end{equation}
For the neutrally buoyant boundary layer with $Ri=0$, the corresponding eigenpair $(\alpha_0,\hat{\bold{q}}_0)$ satisfies:
\begin{equation}\label{eq:residual_wobuo}
\mathcal{R}(\alpha_0,0)\,\hat{\bold{q}}_0=0, \quad \text{with} \; \mathcal{L}_\mathit{q,\mathbf{S}}=0.
\end{equation}
For small but finite $Ri$, only the operator $\mathcal{A}$ in $\mathcal{R}$ \eqref{eq:25} depends on buoyancy effects, while $\mathcal{B}$ and $\mathcal{V}_\mathit{xx}$ remain independent of $Ri$. We therefore expand the operator $\mathcal{A}$ and eigenpair $(\alpha,\hat{\bold{q}})$ in a first-order Taylor series about $Ri=0$ (subscript $0$) as
\begin{subequations}
\begin{gather} \label{eq:Taylor_expansion}
    \mathcal{A}(Ri)=\mathcal{A}_0+\delta\mathcal{A}+\mathcal{O}(Ri^2), \nonumber \\  \alpha(Ri)=\alpha_0+\delta\alpha+\mathcal{O}(Ri^2), \tag{\theequation$a$--$c$} \\ 
    \hat{\bold{q}}(Ri)=\hat{\bold{q}}_0+\delta\hat{\bold{q}}+\mathcal{O}(Ri^2), \nonumber
\end{gather}
\end{subequations}
where the first-order operator perturbation is
\begin{equation}
\delta\mathcal{A}=Ri\underbrace{\left.\dfrac{\partial \mathcal{A}}{\partial Ri}\right|_{\mathit{Ri}=0}}_{\mathcal{C}}, \quad \text{with} \;
\mathcal{C}=\dfrac{\rho^*_\infty}{\Delta\rho^*}\begin{bmatrix} 
0 & 0 & 0 & 0 \\
0 & 0 & 0 & 0 \\
1 & 0 & 0 & 0 \\
0 & 0 & \bar{\rho}-1 & 0 
\end{bmatrix}.
\label{eq:C_term}
\end{equation}
Here $\mathcal{C}$ denotes the first-order buoyancy operator, which is independent of $e_y$. Substituting the Taylor expansions \eqref{eq:Taylor_expansion} and matrix $\mathcal{C}$ \eqref{eq:C_term} into the residual operator in \eqref{eq:residual_wbuo} and collecting terms in powers of $Ri$, the zeroth-order terms recover the neutrally buoyant case, while the first-order terms, $\mathcal{O}(Ri)$, read:
\begin{equation}\label{eq.pert_EVP_O1}
    \mathcal{R}(\alpha_0,0) \,\delta\hat{\bold{q}}=-Ri \, \mathcal{C} \,\hat{\bold{q}}_0+\delta\alpha(\mathcal{B}+2\alpha_0 \, \mathcal{V}_\mathit{xx})\hat{\bold{q}}_0.
\end{equation}
Assuming $\alpha_0$ is a simple (discrete) eigenvalue, let the adjoint eigenvector $\tilde{\bold{q}}_0$ satisfy $\tilde{\bold{q}}^{\dagger}_0 \,\mathcal{R}(\alpha_0,0)=0$, where $(\cdot)^{\dagger}$ represents the Hermitian transpose, with normalisation $\tilde{\bold{q}}^{\dagger}_0 \,(\mathcal{B}+2\alpha_0 \,\mathcal{V}_\mathit{xx})\hat{\bold{q}}_0=1$. At $(\alpha_0,0)$, $\mathcal{R}$ is singular. Thus, to enforce solvability, we project the first-order perturbation equation \eqref{eq.pert_EVP_O1} onto $\tilde{\bold{q}}_0$, yielding the first-order eigenvalue correction
\begin{align}\label{eq.dalpha}
     \delta \alpha =  Ri \, (\tilde{\bold{q}}^{\dagger}_0 \, \mathcal{C} \, \hat{\bold{q}}_0), \quad \text{with} \;  C_0 :=\left.\frac{\partial \alpha}{\partial Ri}\right|_{Ri=0}=\tilde{\bold{q}}^{\dagger}_0 \, \mathcal{C} \, \hat{\bold{q}}_0,
\end{align}
where $C_0$ is the (complex) first-order eigenvalue sensitivity. Note that $C_0$, following the first-order buoyancy operator \eqref{eq:C_term}, can be reformulated as
\begin{equation}
C_0=C_{0,\mathrm{OB}}+C_{0,\mathrm{NOB}},
\label{eq:phys_C0}
\end{equation}
with
\begin{equation}
 C_{0,\mathrm{OB}}=\dfrac{\rho^*_\infty}{\Delta\rho^*}\int_0^{\infty} \tilde{v}_0^{\dagger}\hat{\rho}_0 \, \mathrm{d}y \quad \text{and} \quad   C_{0,\mathrm{NOB}}=\dfrac{\rho^*_\infty}{\Delta\rho^*}\int_0^{\infty} \tilde{T}_0^{\dagger}(\bar{\rho}-1)\hat{v}_0 \, \mathrm{d}y.
 \label{eq:phys_C02}
\end{equation}
The first term $C_{0,\mathrm{OB}}$ represents the classical OB buoyancy coupling, while $C_{0,\mathrm{NOB}}$ accounts for the NOB buoyancy contribution:~density variations in the base flow, through $\bar{\rho}-1$, modulate the coupling between temperature and wall-normal velocity fluctuations. Remarkably, $C_{0,\mathrm{OB}}$ directly governs buoyancy production in the perturbation kinetic-energy budget, linking the buoyancy sensitivity to the energy transfer; see Appendix~\ref{sec:appKE}. Finally, the eigenvalue of the buoyant boundary layer, to order $\mathcal{O}(Ri)$, reads:
\begin{align}\label{eq.dadRi2}
     \alpha = \alpha_0 + C_0 Ri.
\end{align}
The first-order eigenfunction correction $\delta \hat{\bold{q}}$ follows from the first-order perturbation equation \eqref{eq.pert_EVP_O1}, subject to the first-order normalisation constraint $\tilde{\bold{q}}^{\dagger}_0 \,(\mathcal{B}+2\alpha_0 \,\mathcal{V}_\mathit{xx})\,\delta\hat{\bold{q}} + 2 \,\delta \alpha  \,\tilde{\bold{q}}^{\dagger}_0 \mathcal{V}_\mathit{xx} \hat{\bold{q}}_0=0$.

\section{First-order buoyancy correction:~ideal gas}
\label{sec:4}
In this section, we validate and apply the first-order buoyancy correction derived in \S\,\ref{sec:3} to ideal-gas stratified boundary-layer flows. The accuracy of the first-order correction for buoyancy effects on the TS wave is assessed in \S\,\ref{sec:4a} for a nearly incompressible (Blasius) boundary layer. Subsequently, \S\,\ref{sec:4b} examines the parametric variation of the first-order eigenvalue sensitivity $C_0$ across a wide range of Prandtl numbers, Mach numbers, and wall-to-free-stream temperature ratios, with a particular focus on the $N$-factor for transition prediction. 

\subsection{Neutral curves, growth rates, and $N$-factors}
\label{sec:4a}
\begin{figure}
\centering
\includegraphics[angle=-0,trim=0 0 0 0, clip,width=0.85\textwidth]{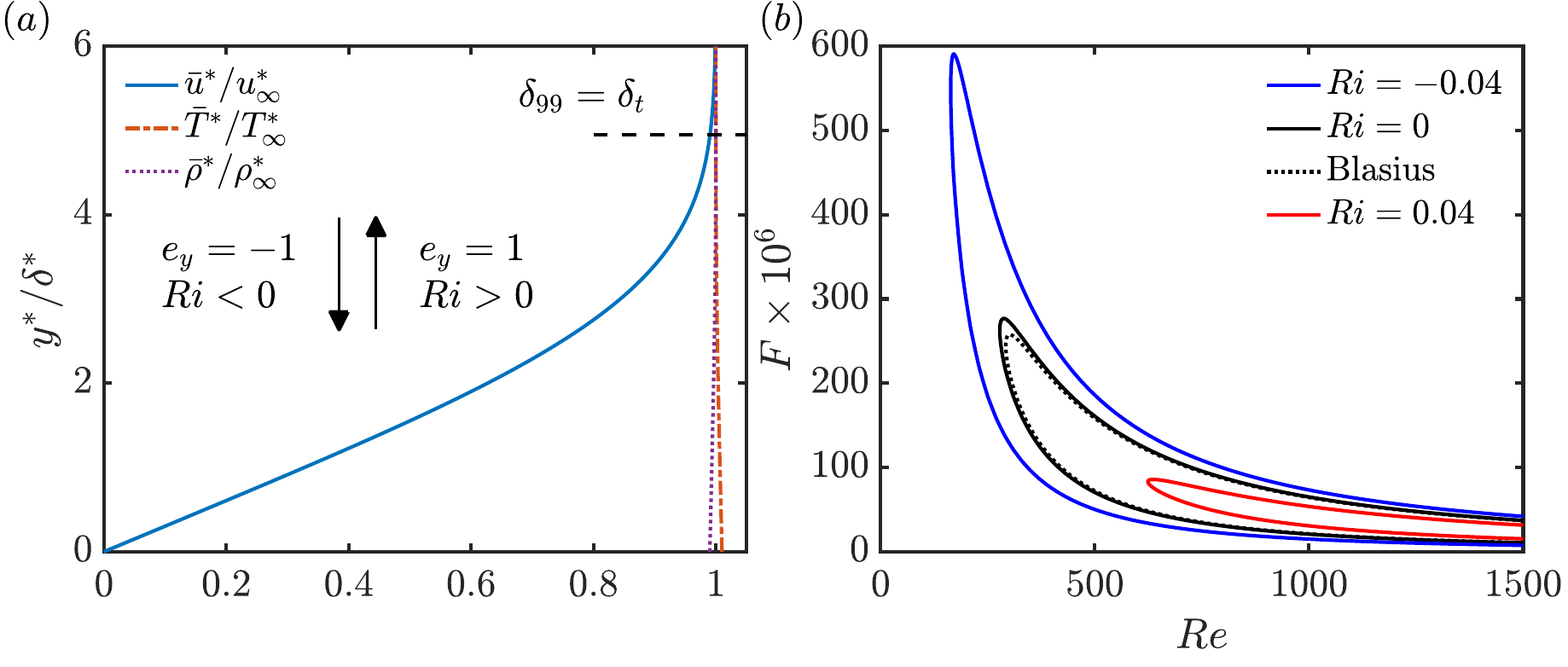}
\captionsetup{justification=justified}
  \caption{(a) Base-flow profiles of streamwise velocity $\bar{u}/u^*_\infty$, temperature $\bar{T}^*/T^*_\infty$, and density $\bar{\rho}^*/\rho^*_\infty$ for $T^*_\mathit{w}/T^*_\infty=1.01$ over the dimensionless wall-normal coordinate $y^*/\delta^*$. The black arrow indicates the direction of gravity. The velocity and thermal boundary-layer thicknesses, $\delta_{99}$ and $\delta_t$, are indicated, respectively. (b) Neutral-stability curves in the $Re$--$F$ plane for $T^*_\mathit{w}/T^*_\infty=1.01$ at $Ri=[-0.04,0,0.04]$. The black dotted line shows the neutral stability of the Blasius profile.  }
\label{fig1}
\end{figure}

We consider a flat-plate boundary layer with free-stream Mach number $M_\infty=0.01$ and $Pr_\infty=1$. A minimal wall-to-free-stream temperature difference (ratio $T^*_\mathit{w}/T^*_\infty=1.01$, i.e.~$|\Delta\rho^*/\rho^*_\infty|=0.01$) introduces mild stratification while keeping the flow nearly incompressible. The reference case corresponds to the neutrally buoyant boundary layer with source-term vector $\mathbf{S}=0$ (see \eqref{eq:source2}). The base flow is shown in figure~\ref{fig1}(a), and the corresponding neutral-stability curve ($\Im\{\alpha_0\}=0$) in the $Re$--$F$ plane is presented in figure~\ref{fig1}(b), where $F=\omega^*\mu^*_\infty/(\rho^*_\infty u^{*2}_\infty)$ denotes the dimensionless frequency. 
Stable ($Ri>0$) and unstable ($Ri<0$) stratification use the same base flow (figure~\ref{fig1}a) but opposite gravity directions, resulting in a different Richardson number $Ri$ (see \eqref{eq:Ri}). Flipping the gravity direction changes the sign of $Ri$ without modifying the base flow, thereby isolating the buoyancy effect on instability. Positive $Ri$ stabilises the TS wave as reported by \citet{Parente2020}, whereas $Ri<0$ promotes instability. The neutral curve for the neutrally buoyant boundary layer ($Ri=0$) slightly deviates from that of the incompressible (Blasius) boundary layer due to non-uniform temperature-dependent fluid viscosity (see also \citet{Wall1997}).

\begin{figure}
\centering
\includegraphics[angle=-0,trim=0 0 0 0, clip,width=0.88\textwidth]{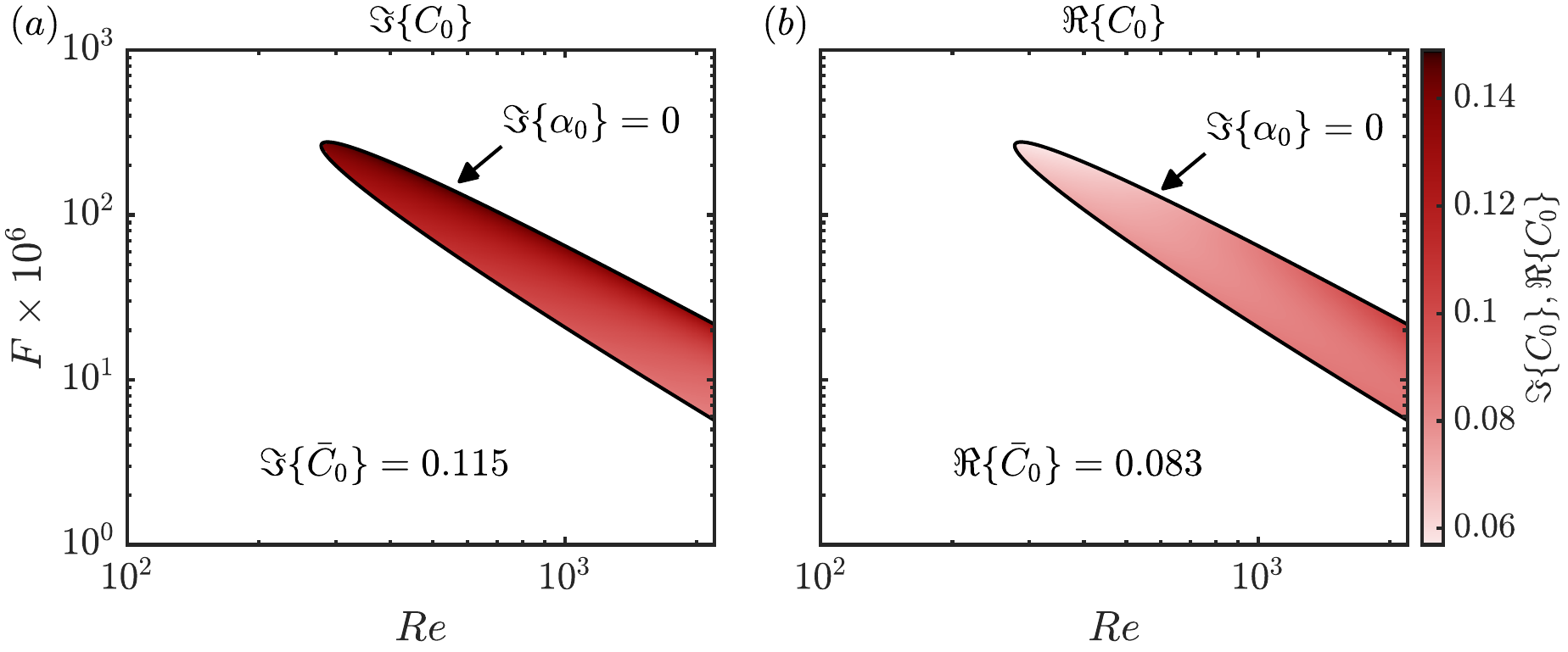}
\captionsetup{justification=justified}
  \caption{Contours of (a) $\Im\{C_0\}$ and (b) $\Re\{C_0\}$ in the $Re$--$F$ plane. The black solid line indicates the neutral-stability curve where $\Im\{\alpha_0\}=0$.}
\label{fig2}
\end{figure}

The sensitivity of the first-order buoyancy correction \eqref{eq.dalpha} is governed by $C_0$. Figure~\ref{fig2} displays its imaginary ($\Im\{C_0\}$) and real ($\Re\{C_0\}$) parts in panels (a) and (b), respectively. 
We evaluate $C_0$ at each $(Re, F)$. Because $C_0$ is computed for the neutrally buoyant case (subscript $0$, $Ri=0$), it is independent of the stratification level; within the neutral-stability curve its contours vary only weakly with $Re$ and $F$, especially $\Re\{C_0\}$. Consequently, it is not necessary to re-solve the eigenvalue problem \eqref{eq:residual_wobuo} at every $(Re, F)$ to apply the
first-order correction. Instead, we adopt a single
representative value $\bar{C}_0$, defined as the arithmetic mean of $C_0$ inside the neutral curve, i.e.~$\bar{C}_0 = 1/A \iint_{{U}_\mathit{\alpha}} C_0(Re,F) \, \mathrm{d}Re \, \mathrm{d}F$, where $U_\mathit{\alpha}$ denotes the unstable region enclosed by the neutral curve and $A$ is its area. Substituting $\bar{C}_0$ into the first-order correction \eqref{eq.dadRi2} corrects the stability of the neutrally buoyant reference case by the factor $\bar{C}_0 Ri$. Note that in this weakly stratified case (figure~\ref{fig1}a), the NOB contribution in the buoyancy sensitivity \eqref{eq:phys_C0} is absent, thus $C_0 \approx C_{0,\mathrm{OB}}$.

\begin{figure}
\centering
\includegraphics[angle=-0,trim=0 0 0 0, clip,width=0.7\textwidth]{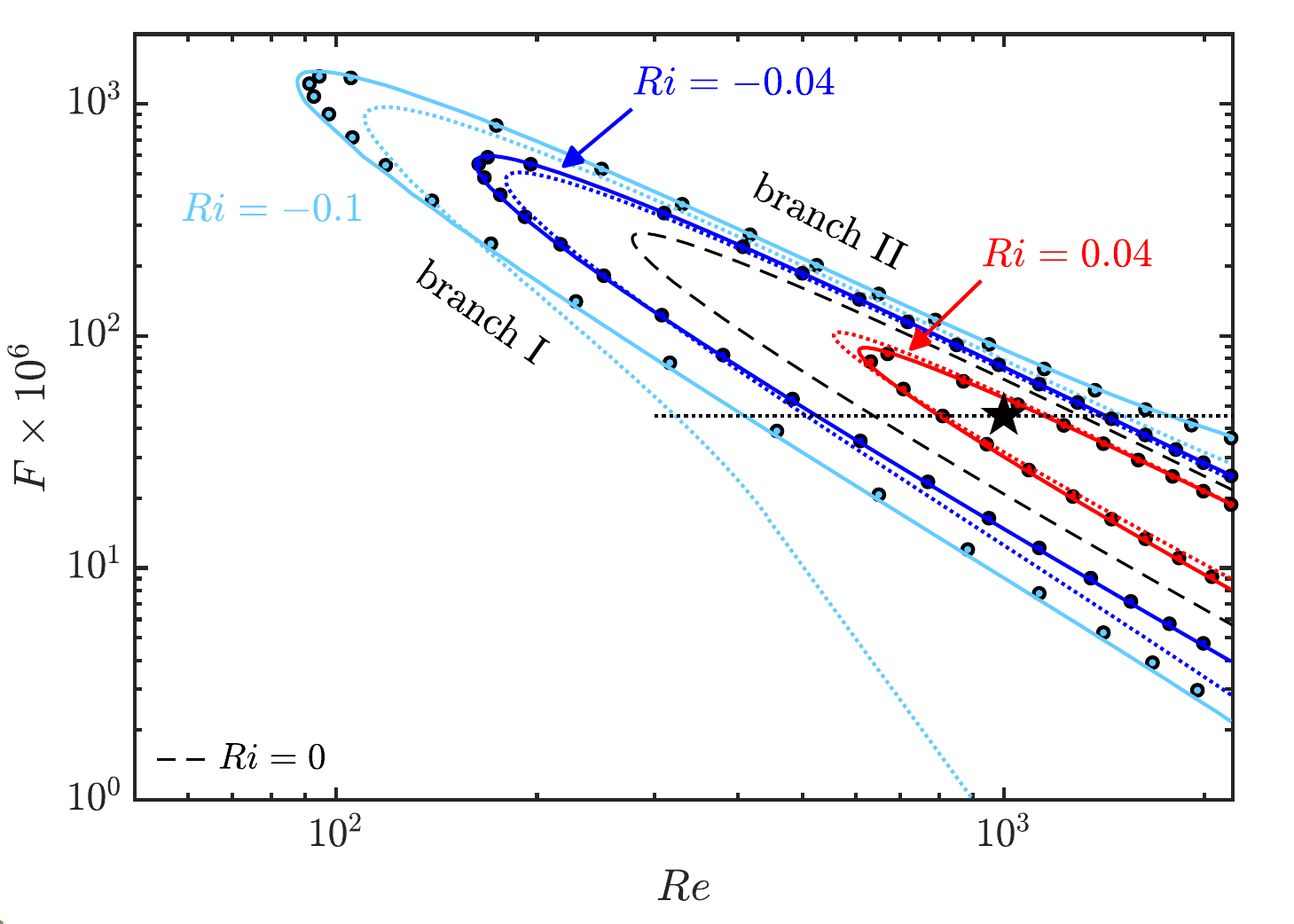}
\captionsetup{justification=justified}
  \caption{Neutral-stability curves in the $Re$--$F$ plane for stably and unstably stratified cases at $Ri=[-0.1,-0.04,0.04]$. Symbols (\textcolor{black}{$\circ$}) show results from the buoyant eigenvalue problem \eqref{eq:residual_wbuo}; solid lines denote the first-order correction \eqref{eq.dadRi2} evaluated with $C_0$; dotted lines depict the first-order correction \eqref{eq.dadRi2} evaluated with $\bar{C}_0$. The neutrally buoyant case at $Ri=0$ is indicated with a black dashed line. The black pentagram indicates the location at which the eigenfunctions are extracted in Appendix~\ref{sec:appB}. }
\label{fig3}
\end{figure}

We now apply the first-order buoyancy correction \eqref{eq.dadRi2}, using both $C_0$ and $\bar{C}_0$, to the neutrally buoyant reference case and, for validation, compare against the buoyant (direct) eigenvalue problem \eqref{eq:residual_wbuo} for unstably ($Ri=[-0.1,-0.04]$) and stably ($Ri=0.04$) stratified cases. We emphasise that \eqref{eq.dadRi2} requires only a single calculation at $Ri=0$ and is then used for moderate $Ri$. For instance, evaluating \eqref{eq.dadRi2} requires only $\sim 0.1\%$ of the wall-clock time needed to solve the buoyant eigenvalue problem \eqref{eq:residual_wbuo} with $N_\mathit{y}=201$ wall-normal collocation points. In figure~\ref{fig3}, the neutral stability is displayed in the $Re$--$F$ plane on double-logarithmic axes. Circles denote solutions of the buoyant eigenvalue problem \eqref{eq:residual_wbuo}, while solid and dotted lines show the first-order correction \eqref{eq.dadRi2} using the local $C_0$ and $\bar{C}_0$, respectively. 
Using $C_0$, excellent agreement of branches I and II is obtained up to $Ri=-0.1$. The averaged $\bar{C}_0$ continues to provide a robust prediction of the neutral curve up to $Ri=-0.04$; however, for $Ri=-0.1$ the discrepancies in the critical Reynolds number and along branches I and II at low frequencies can exceed $20\%$. Accordingly, $\bar{C}_0$ remains valid for weak stratification, whereas for stronger buoyancy effects, nonlinear deviations become significant, and the local $C_0$ is required for accurate stability predictions. A comparison of the corresponding eigenfunctions between the buoyant eigenvalue problem \eqref{eq:residual_wbuo} and the first-order correction \eqref{eq.dadRi2}, extracted within the unstable region (indicated by the black pentagram in figure~\ref{fig3}), is reported in Appendix~\ref{sec:appB}.

\begin{figure}
\centering
\includegraphics[angle=-0,trim=0 0 0 0, clip,width=0.85\textwidth]{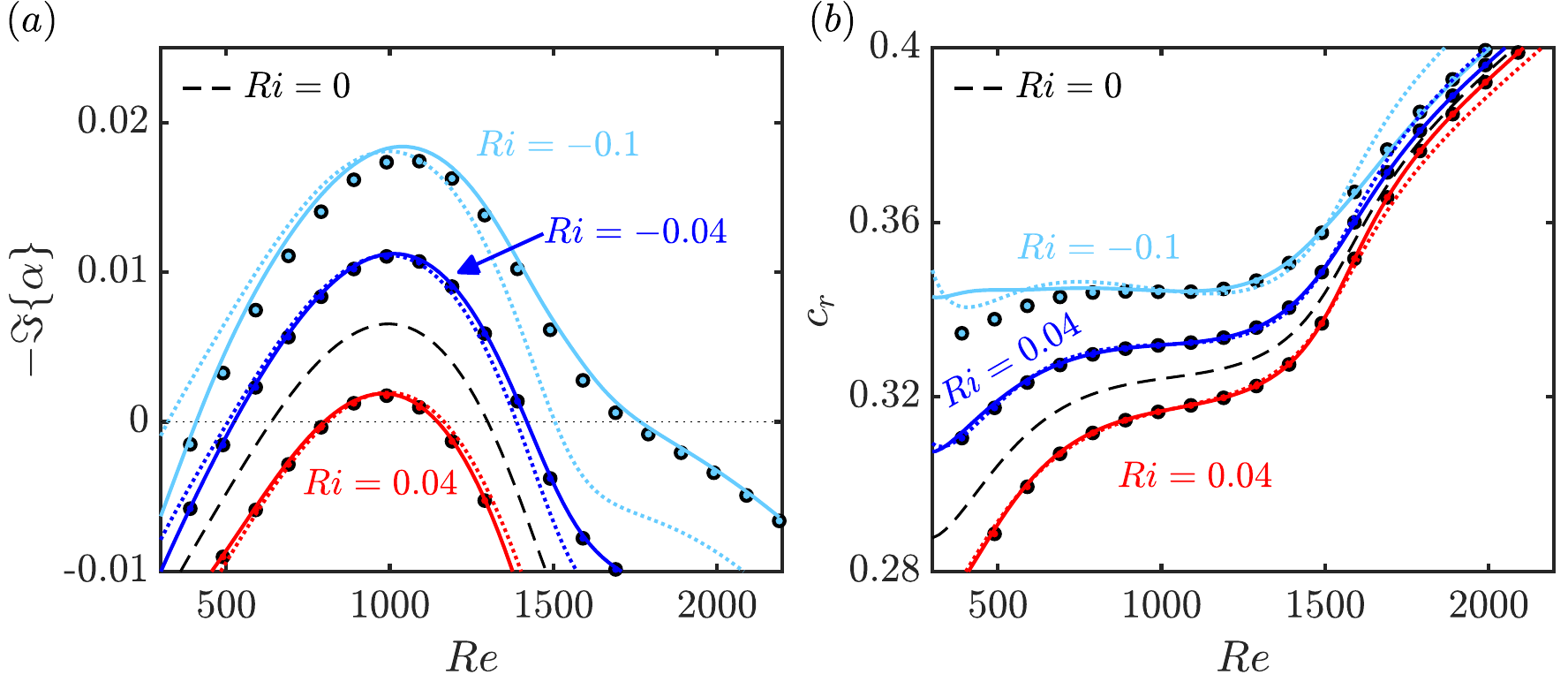}
\captionsetup{justification=justified}
  \caption{(a) Growth rate and (b) phase speed as functions of $Re$ at $F=45 \times 10^{-6}$ for stably and unstably stratified cases at $Ri=[-0.1,-0.04,0.04]$. Symbols (\textcolor{black}{$\circ$}) show results from the buoyant eigenvalue problem \eqref{eq:residual_wbuo}; solid lines denote the first-order correction \eqref{eq.dadRi2} evaluated with $C_0$; dotted lines depict the first-order correction \eqref{eq.dadRi2} evaluated with $\bar{C}_0$. The neutrally buoyant case at $Ri=0$ is indicated with a black dashed line.}
\label{fig4}
\end{figure}

Fixing the frequency at $F=45 \times 10^{-6}$, figure~\ref{fig4} compares the evolution of the growth rate $\Im\{\alpha\}$ (panel a) and the real part of the phase speed $c_\mathit{r}=\omega/\Re\{\alpha\}$ (panel b) as functions of $Re$.
Unstable stratification at $Ri=-0.1$ yields a maximum growth rate approximately three times larger than that of the neutrally buoyant case $\Im\{\alpha=\alpha_0\}$ (black dashed line). Conversely, stable stratification leads to a reduction in phase speed. Using $C_0$, the most unstable case at $Ri=-0.1$ is predicted with relative errors of approximately $5\%$ at $\max\{-\Im\{\alpha\}\}$ and $4\%$ in the critical Reynolds number. The largest discrepancies between $C_0$ and $\bar{C}_0$ occur near $\Im\{\alpha\}=0$, consistent with figure~\ref{fig3}. Notably, for $Ri=-0.1$, the maximum growth rate at $Re\approx1000$ is accurately captured when using $\bar{C}_0$, with a relative error similar to that obtained with $C_0$. A comparable trend in the relative error is seen for $c_\mathit{r}$ in figure~\ref{fig4}(b). Under stronger unstable stratification, the phase speed is well predicted in the vicinity of the largest growth rate $\max\{-\Im\{\alpha\}\}$. For weaker stratification, both first-order corrections based on $C_0$ and $\bar{C}_0$ accurately predict the evolution of $c_\mathit{r}$.

\begin{figure}
\centering
\includegraphics[angle=-0,trim=0 0 0 0, clip,width=0.57\textwidth]{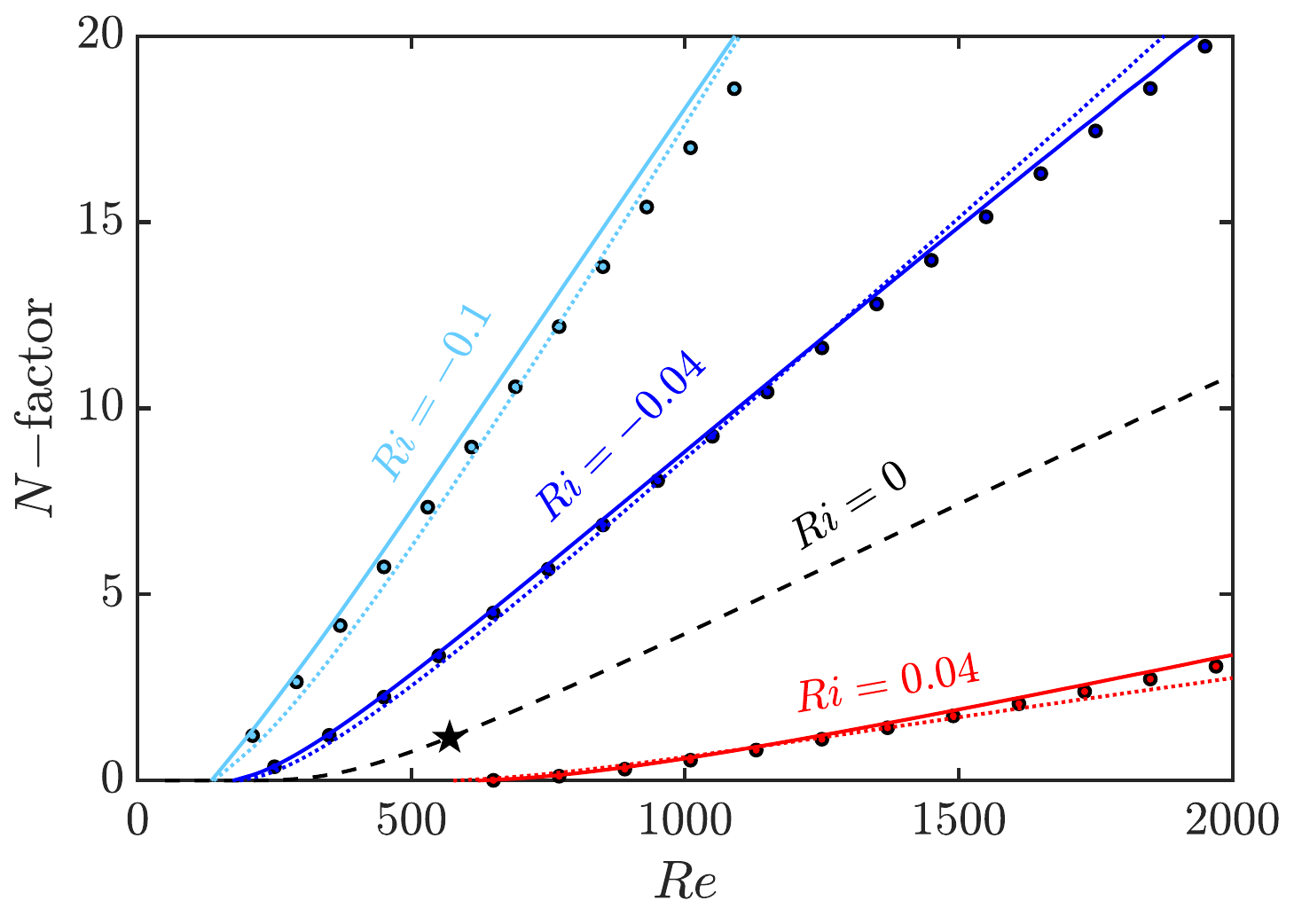}
\captionsetup{justification=justified}
  \caption{$N$-factor envelopes at $Ri=[-0.1,-0.04,0,0.04]$. Symbols (\textcolor{black}{$\circ$}) show results from the buoyant eigenvalue problem \eqref{eq:residual_wbuo}; solid lines denote the first-order correction \eqref{eq.dadRi2} evaluated with $C_0$; dotted lines depict the first-order correction \eqref{eq.dadRi2} evaluated with $\bar{C}_0$. The $N$-factor of the neutrally buoyant case is indicated with a black dashed line. The location of maximum amplification rate, $\max\{-\Im\{\alpha_0\}/Re\}$, is indicated with a black pentagram.  }
\label{fig5}
\end{figure}

In addition to the local amplification shown in figure~\ref{fig4}(a), the calculation of the $N$-factor is essential for practical transition prediction. Using the first-order buoyancy correction \eqref{eq.dadRi2}, the $N$-factor, $N = -\int_{x_0}^{x}\Im\{\alpha\}\,\mathrm{d}x$ for unstably and stably stratified flows can be written as:
\begin{equation}\label{eq:Nfactor}
    N=N_\mathit{Ri=0}+ Ri\,
    \left.\frac{\partial N}{\partial Ri}\right|_\mathit{Ri=0}= -\int_{x_0}^{x}\Im\{\alpha_0\}\,\mathrm{d}x
-Ri\int_{x_0}^{x}\Im\{C_0\}\,\mathrm{d}x.
\end{equation}
Figure~\ref{fig5} displays the $N$-factor envelopes obtained from \eqref{eq:Nfactor} using both $C_0$ and $\bar{C}_0$, compared with the solution of the buoyant eigenvalue \eqref{eq:residual_wbuo} for different stratification levels. Note that the $N$-factor envelope represents, at each streamwise position $x$, the maximum $N$-factor over all frequencies $F$.
Unstable stratification yields significantly higher $N$-factors, consistent with the enhanced growth rates in figure~\ref{fig4}(a), leading to earlier transition, while stable stratification delays transition. Note that $N$-factor predictions do not explicitly consider the role of the initial disturbance amplitude and therefore require calibration with experimental data, which are not yet available for the present stratified flow conditions. Overall, the first-order correction reproduces the $N$-factor well, with the most accurate prediction obtained when using $C_0$ rather than $\bar{C}_0$. The latter is observed to be accurate for weak stratification near the maximum of the amplification rate, defined as $\max\{-\Im\{\alpha_0\}/Re\}$ (neutrally buoyant case) and indicated by a pentagram in figure~\ref{fig5} following \citet{Mack1984}. This low-$Re$ amplification maximum is typical of TS modes and, due to the integral nature of the $N$-factor, the region around this peak dominates the integral amplification. Consequently, evaluating $C_0$ at the location of $\max\{-\Im\{\alpha_0\}/Re\}$ captures the dominant buoyancy contribution to the $N$-factor envelope. 
Consistent with this interpretation, the buoyancy sensitivity 
$\partial N/\partial Ri|_\mathit{Ri=0}$ is largest at low $Re$. Therefore, $N$-factor envelopes obtained by evaluating $C_0$ at $\max\{-\Im\{\alpha_0\}/Re\}$ agree well with the actual $N$-factor (buoyant eigenvalue problem) near the amplification maximum (not shown) but become less accurate farther downstream, where the averaged $\bar{C}_0$ performs better.

\subsection{Base-flow variation: buoyancy sensitivity}
\label{sec:4b}
\begin{figure}
\centering
\includegraphics[angle=-0,trim=0 0 0 0, clip,width=0.88\textwidth]{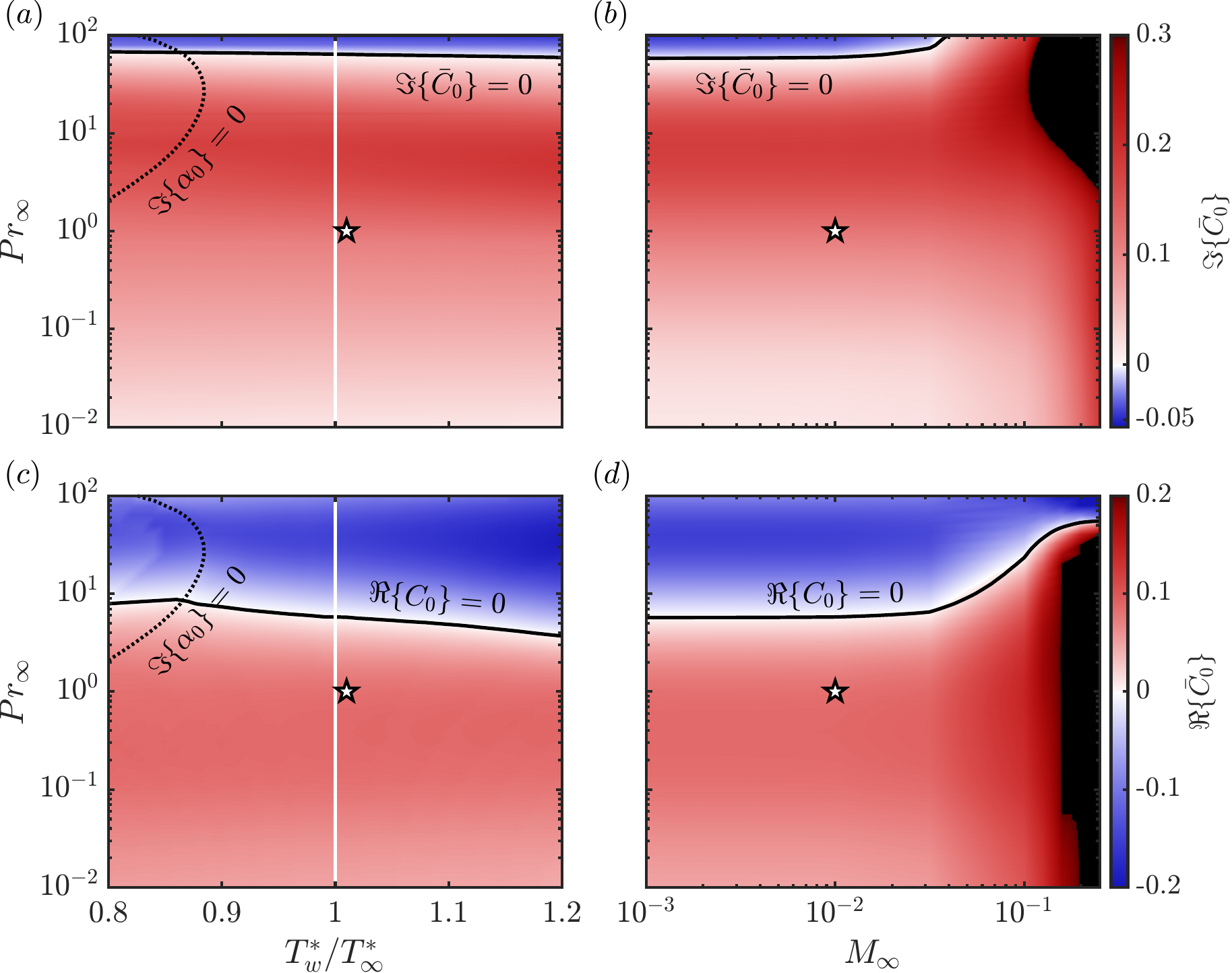}
\captionsetup{justification=justified}
  \caption{Contours of (a,b) $\Im\{\bar{C}_0\}$ and (c,d) $\Re\{\bar{C}_0\}$ in the (a,c) $T^*_\mathit{w}/T^*_\mathit{\infty}$--$Pr_\mathit{\infty}$ plane at $M_\mathit{\infty}=0.01$ and in the (b,d) $M_\infty$--$Pr_\infty$ plane at $T^*_\mathit{w}/T^*_\mathit{\infty}=1.01$. In panels (a,c), the region enclosed by the black dashed line $\Im\{\alpha_0\}=0$ corresponds to $\max\{-\Im\{\alpha_0\}\}<0$ (stable TS mode). The white star indicates the base-flow conditions of \S~\ref{sec:4a}. In panels (a,c), when $T^*_\mathit{w}/T^*_\mathit{\infty}=1$, $Ri=0$. The black solid line indicates $\Im\{\bar{C}_0\}=0$ in panels (a,b) and $\Re\{\bar{C}_0\}=0$ in panels (c,d).}
\label{fig6}
\end{figure}

After verifying the first-order buoyancy correction on a representative base flow, we examine its behaviour across different flow parameters, i.e.~Mach number $M_\infty$, Prandtl number $Pr_\infty$, and wall-to-free-stream temperature ratios $T^*_\mathit{w}/T^*_\infty$, which modify the base-flow density profile and hence the buoyancy variation in $\delta\mathcal{A}$ (see the first-order buoyancy operator \eqref{eq:C_term}). Figures~\ref{fig6}(a,b) show $\Im\{\bar{C}_0\}$, with $\bar{C}_0$ computed analogously to figure~\ref{fig3}. Note that the black dashed line marks $\Im\{\alpha_0\}=0$; thus, in that region (at large $Pr_\infty$ with $T^*_\mathit{w}/T^*_\mathit{\infty} \lesssim 0.88$), no unstable $\alpha_0$ exists in the $(Re,F)$ plane up to $Re=2000$. In such cases, $C_0$ is evaluated locally for the stable TS mode. 
In panel (a) at $M_\infty=0.01$, $\Im\{\bar{C}_0\}$ is strongly dependent on $Pr_\infty$. After peaking near $Pr_\infty \simeq 7$ (i.e.~water), the buoyancy sensitivity decreases monotonically regardless of $T^*_\mathit{w}/T^*_\infty$ and eventually for $Pr_\infty\gtrsim70$ (e.g.~oil) changes sign. This sign reversal indicates that the role of buoyancy on modal stability switches at high $Pr_\infty$: for a given sign of $Ri$, flow configurations that are destabilising at low $Pr_\infty$ become stabilising at high $Pr_\infty$, and vice versa. This non-monotonic dependence of $\Im\{\bar{C}_0\}$ on $Pr_\infty$ reflects the mechanism identified by \citet{Parente2020}: increasing $Pr_\infty$ reduces thermal diffusion, altering the phase between density and velocity perturbations -- stabilising the flow through buoyancy effects at low $Pr_\infty$ and destabilising it at high $Pr_\infty$. In contrast, the sensitivity to variations in $T^*_\mathit{w}/T^*_\infty$ is small. A similar behaviour holds in figure~\ref{fig6}(b) for $M_\infty\lesssim 0.1$; at higher $M_\infty$, compressibility increases $\Im\{\bar{C}_0\}$, enhancing TS growth under unstable stratification. Nevertheless, buoyancy effects remain most significant in the low-Mach regime, where compressibility effects are weak and stratification effects are most pronounced.

Figures~\ref{fig6}(c,d) show the contours of the real part of the buoyancy sensitivity, $\Re\{\bar{C}_0\}$. This quantity governs the buoyancy-induced phase-speed modification, $\delta c_\mathit{r}=-\omega\Re\{C_0\}Ri/\Re\{\alpha_0\}^2$, and determines the net buoyancy-production term in the perturbation kinetic-energy budget (see Appendix~\ref{sec:appKE}). The dependence of $\Re\{\bar{C}_0\}$ on $T^*_\mathit{w}/T^*_\infty$ (panel c) and $M_\infty$ (panel d) is similar to that observed in figures~\ref{fig6}(a,b), but a distinct behaviour emerges with increasing $Pr_\infty$. The locus $\Re\{\bar{C}_0\}=0$ lies near $Pr_\infty\approx\mathcal{O}(10)$, indicating the region where buoyancy switches from decreasing to increasing phase speed. At low $Pr_\infty$, where thermal diffusion dominates, buoyancy decreases the phase speed for stable stratification ($Ri>0$), whereas at high $Pr_\infty$ where $\Re\{\bar{C}_0\}<0$, buoyancy increases the phase speed under stable stratification. Regarding the buoyancy–production term $\mathcal{P}_{\mathbf{S}}$ in Appendix~\ref{sec:appKE}, the sign change of $\Re\{\bar{C}_0\}$ at high $Pr_\infty$ causes the net buoyancy production \eqref{eq:Prod_C0_Ri} to become negative. Consequently, for unstable stratification ($Ri<0$), we have $\Re\{\mathcal{P}_\mathbf{S}\}<0$, indicating that buoyancy has a stabilising effect on the perturbation kinetic-energy budget \eqref{eq:KE}. However, figures~\ref{fig6}(a,b) show that $\Im\{\bar{C}_0\}$ is higher around $Pr_\infty=\mathcal{O}(10)$ than at $Pr_\infty=\mathcal{O}(1)$. This enhanced destabilisation arises ultimately from the larger thermodynamic term $\mathcal{T}$ in perturbation kinetic-energy budget \eqref{eq:KE} (not shown here), which is driven by stronger temperature fluctuations associated with the thinner thermal boundary layer, in agreement with \citet{Parente2020}.

\begin{figure}
\centering
\includegraphics[angle=-0,trim=0 0 0 0, clip,width=0.88\textwidth]{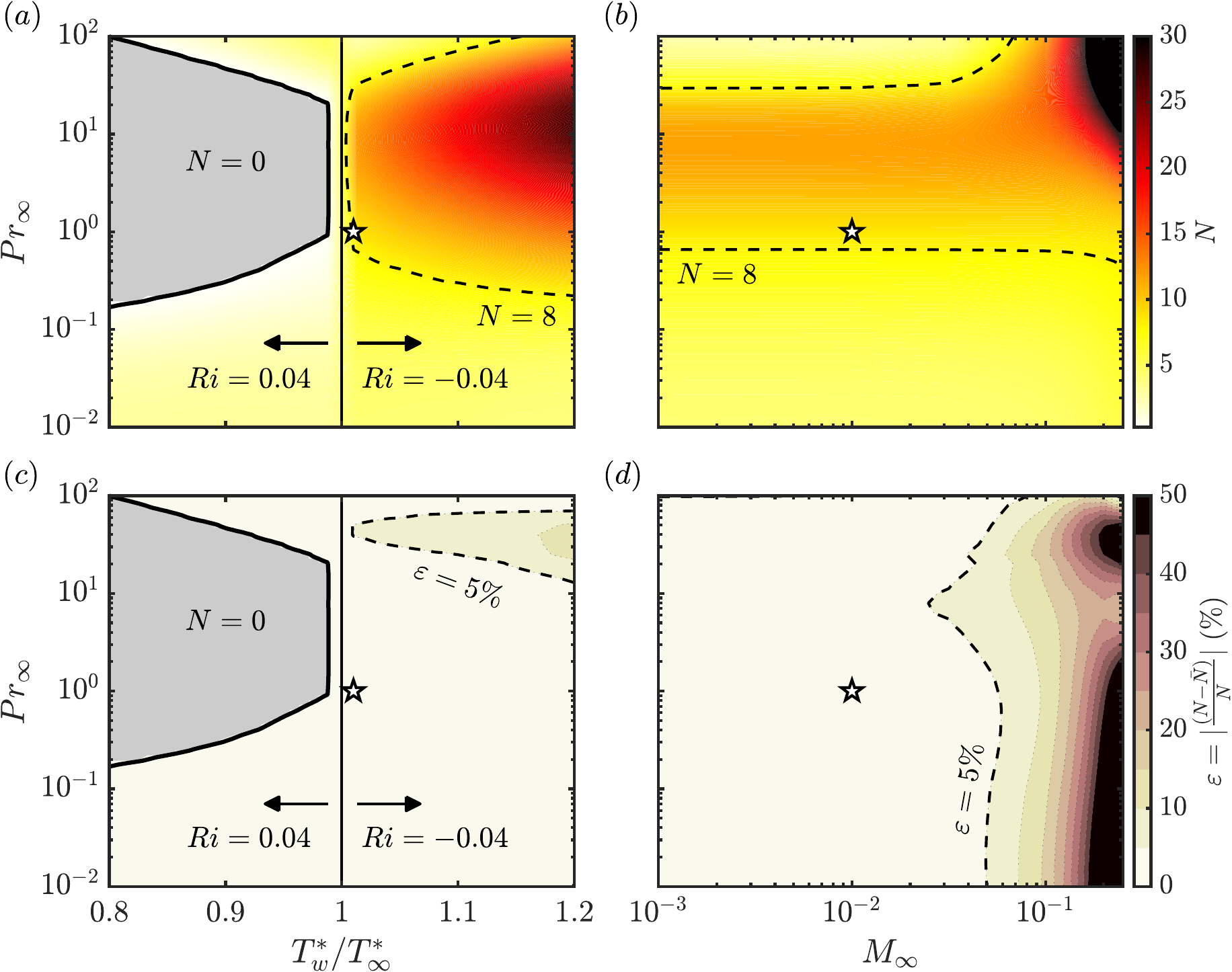}
\captionsetup{justification=justified}
  \caption{Contours of (a,b) $N$-factor envelope and (c,d) relative error $\varepsilon=|(N-\bar{N})/N|$ (in percentage), evaluated at $Re=1000$. Panels (a,c) show the $T^*_\mathit{w}/T^*_\mathit{\infty}$--$Pr_\mathit{\infty}$ plane at $M_\mathit{\infty}=0.01$; panels (b,d) show the $M_\infty$--$Pr_\infty$ plane at $T^*_\mathit{w}/T^*_\mathit{\infty}=1.01$. We use $|Ri|=0.04$ with gravity directed toward the wall ($e_y=-1$). In panels (a,c), the grey region corresponds to $N=0$; note that $Ri=0$ when $T^*_\mathit{w}/T^*_\mathit{\infty}=1$. The white star indicates the base-flow conditions of \S\,\ref{sec:4a}. The black dashed line indicates $N=8$ in panels (a,b).}
\label{fig7}
\end{figure}
Figures~\ref{fig7}(a,b) show contours of the $N$-factor envelope at $Re=1000$, computed as $N=\int^x_{x_0}-\Im\{\alpha\}\,\mathrm{d}x$ from the buoyant eigenvalue problem \eqref{eq:residual_wbuo}. Panel (a) presents the results in the $T^*_\mathit{w}/T^*_\infty$--$Pr_\infty$ plane at $M_\infty=0.01$, while panel (b) shows the $M_\infty$--$Pr_\infty$ plane at $T^*_\mathit{w}/T^*_\infty=1.01$. We set $|Ri|=0.04$ with gravity directed toward the wall ($e_y=-1$). For wall cooling ($T^*_\mathit{w}/T^*_\infty<1$) the colder, denser fluid lies at the wall, yielding stable stratification $Ri>0$. Conversely, wall heating ($T^*_\mathit{w}/T^*_\infty>1$) produces unstable stratification ($Ri<0$) with hotter, lighter fluid at the wall. For $T^*_\mathit{w}/T^*_\infty<1$, wall cooling stabilises, as expected, the TS mode (see also $\Im\{\alpha_0\}=0$ in figure~\ref{fig6}a), and this effect is further enhanced by stable stratification ($Ri=0.04$), leading to an extended region of complete stabilisation ($N=0$). In comparison, for a neutrally buoyant case, the wall-to-free-stream temperature ratio required to achieve $N=0$ would be $T^*_\mathit{w}/T^*_\infty \approx 0.92$ at $Pr_\infty\approx 10$. For $T^*_\mathit{w}/T^*_\infty>1$, unstable stratification ($Ri=-0.04$) substantially increases the $N$-factor, particularly at $Pr_\infty=\mathcal{O}(1)$, where the buoyancy sensitivity peaks (see figure~\ref{fig6}a). For $T^*_\mathit{w}/T^*_\infty=1.2$ and $Pr_\infty \approx 10$, the neutrally buoyant boundary layer would exhibit an $N$-factor approximately $40\%$ lower. Panel (b) shows that $N$ is nearly insensitive to $M_\infty$ in the low-Mach-number regime ($M_\infty \lesssim 0.1$), but exceeds $40$ in the high-$Pr_\infty$, moderate-$M_\infty$ region. For instance, at $M_\infty=0.2$ and $Pr_\infty \approx 100$, the unstably stratified boundary layer ($Ri=-0.04$) reaches an $N$-factor almost eight times larger than the neutrally buoyant case. 

The accuracy of the first-order buoyancy correction when using $\bar{C}_0$ is assessed through the relative $N$-factor error $\varepsilon=|(N-\bar{N})/N|$, shown in figures~\ref{fig7}(c,d). The $N$-factors are extracted from the corresponding envelope at $Re=1000$, where $N$ denotes the $N$-factor from the buoyant eigenvalue problem \eqref{eq:residual_wbuo} (cf.~figures~\ref{fig7}(a,b)) and $\bar{N}$ is computed using $\bar{C}_0$ via the first-order correction of the $N$-factor \eqref{eq:Nfactor}. Across the entire $Pr_\infty$ and $T^*_\mathit{w}/T^*_\infty$ ranges considered, the $\bar{C}_0$-based first-order buoyancy correction reproduces the integral amplification well, with typical errors below $5\%$. However, as compressibility increases ($M_\infty \gtrsim 0.1$), $\varepsilon$ rises substantially, exceeding $50\%$ in some regions. This indicates that higher-order buoyancy effects become non-negligible at moderate Mach numbers, such that the first-order perturbation theory itself becomes insufficient, and the buoyant eigenvalue problem \eqref{eq:residual_wbuo} must be solved at each $Ri$.

\section{First-order buoyancy correction:~supercritical fluids}
\label{sec:5}

Beyond the ideal-gas regime explored in \S\,\ref{sec:4}, we apply the first-order perturbation theory to the non-ideal gas regime of fluids at supercritical pressure. In the vicinity of the critical point, strong density variations ($\Delta \rho^*/\rho^*_\infty \approx \mathcal{O}(1)$) across the boundary layer become extremely important \citep{Wang2023,Draskic2025}. Modal instabilities in boundary layers with supercritical fluids have been widely investigated \citep{Ren2019b,Robinet2019}, with recent work identifying the crucial role of baroclinic effects in the presence of large density variations \citep{Bugeat2024,Boldini2025b}. However, all stability analyses thus far have neglected gravitational effects. Since buoyancy can profoundly 
affect the modal stability of ideal-gas boundary layers (see \S\,\ref{sec:4}) and density stratification in supercritical fluids is orders of magnitude larger, how buoyancy effects alter the instability mechanisms in boundary layers with supercritical fluids remains an open question. Here, we address this by systematically isolating buoyancy contributions through the first-order buoyancy correction.

\begin{figure}
\centering
\includegraphics[angle=-0,trim=0 0 0 0, clip,width=0.85\textwidth]{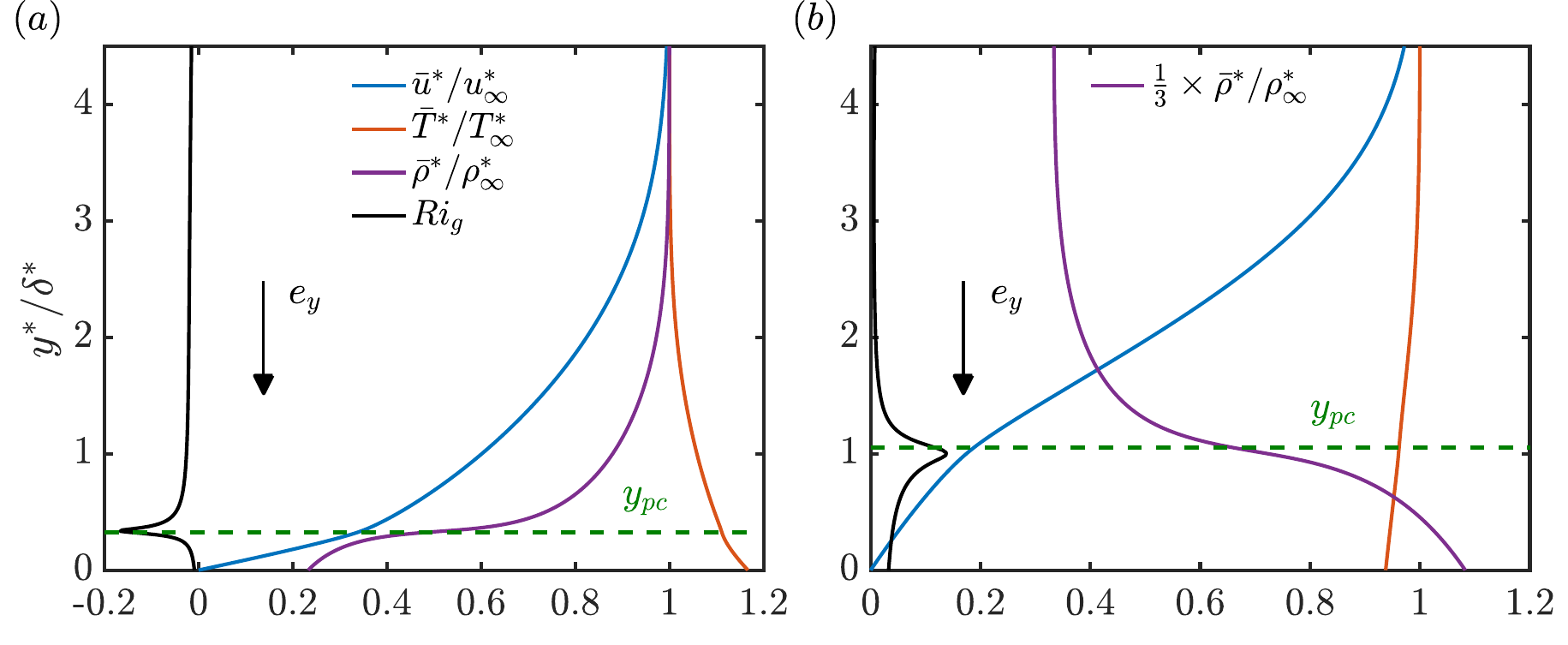}
\captionsetup{justification=justified}
  \caption{Base-flow profiles of streamwise velocity $\bar{u}^*/u^*_\infty$, temperature $\bar{T}^*/T^*_\infty$, density $\bar{\rho}^*/\rho^*_\infty$, and local gradient Richardson number $Ri_\mathit{g}$ over the dimensionless wall-normal coordinate $y^*/\delta^*$. (a) Wall-heating. (b) Wall-cooling. The green dashed line at $y=y_\mathit{pc}$ indicates the location where $\bar{T}^*=T^*_\mathit{pc}$. The black arrow indicates the direction of gravity.}
\label{fig8}
\end{figure}

We consider two boundary layers with carbon dioxide ($\text{CO}_2$) at a supercritical pressure of $\SI{80}{bar}$ (critical pressure of $\SI{73.9}{bar}$) under pseudo-boiling conditions \citep{Banuti2015}, where the boundary-layer temperature crosses the pseudo-critical temperature $T^*_\mathit{pc}$ at which the specific isobaric heat capacity reaches its maximum. The first case features wall heating with $T^*_\infty/T^*_\mathit{pc}=0.90$ and $T^*_\mathit{w}/T^*_\mathit{pc}=1.05$ at $M_\infty=0.05$, conditions known to trigger an inflection inviscid instability (Mode II) \citep{Ren2019b}. The second case involves wall cooling following the study of \citet{Ren2025} with $T^*_\infty/T^*_\mathit{pc}=1.04$ and $T^*_\mathit{w}/T^*_\mathit{pc}=0.975$ at $M_\infty=0.2$, which produces a similar inflectional inviscid mode. Base-flow profiles of streamwise velocity $\bar{u}^*/u^*_\infty$, temperature $\bar{T}^*/T^*_\infty$, density $\bar{\rho}^*/\rho^*_\infty$, and local gradient Richardson number $Ri_\mathit{g} = -g^* (\mathrm{d}\bar{\rho}^*/\mathrm{d}y^*)/ [\rho^*(\mathrm{d}\bar{u}^*/\mathrm{d}y^*)^2]$ are shown in figure~\ref{fig8} for wall heating (panel a) and wall cooling (panel b).
In the wall-heating case at $M_\infty=0.05$, the temperature profile crosses $T^*_\mathit{pc}$ at $y=y_\mathit{pc}$ (pseudo-critical point with continuous fluid property change), generating a near-wall vapour-like region with lower density than the free stream  ($|\Delta \rho^*/\rho^*_\infty| \approx 0.77$). This flow configuration produces unstable stratification with $Ri=-0.037$ (corresponding to $Fr=4.54$) and a negative $Ri_\mathit{g}$ that reaches its minimum at $y_\mathit{pc}$, where gradients of thermophysical properties are strongest. Conversely, the wall-cooling case creates a dense liquid-like region near the wall ($|\Delta \rho^*/\rho^*_\infty| \approx 2.21$), yielding stable stratification with $Ri=0.058$ (corresponding to $Fr=6.22$) and a positive $Ri_\mathit{g}$ throughout the boundary layer (maximum at $y_\mathit{pc}$). Thus, the large density gradients at the pseudo-critical point $y_\mathit{pc}$ intensify buoyancy effects, which either stabilise the flow (when $Ri_\mathit{g}>0$, stable stratification) or further amplify the underlying modal instability (when $Ri_\mathit{g}<0$, unstable stratification).

We examine the linear stability of the wall-heating case in figure~\ref{fig9}; the wall-cooling case is presented in Appendix~\ref{sec:appD}. For both configurations, the most amplified disturbances are two-dimensional \citep{Boldini2024}, justifying the two-dimensional framework adopted for the first-order buoyancy correction in \S~\ref{sec:3}. Panel (a) shows neutral-stability curves in the $Re$--$F$ plane for two levels of unstable stratification: $Ri=-0.037$, corresponding to the base flow in figure~\ref{fig8}(a), and $Ri=-0.1$, obtained using the same thermodynamic conditions but at $M_\infty=0.03$. Unstable stratification shifts the neutral curve toward lower $Re$, promoting earlier instability. The background contours of $\Im\{C_0\}$ (panel a) vary strongly within the neutral curve, reflecting the sharp pseudo-boiling-induced gradients that locally modulate buoyancy sensitivity. Consequently, the arithmetic mean $\Im\{\bar{C}_0\}$, as in figure~\ref{fig2}, is not representative under such strong variable-property conditions; the first-order buoyancy correction \eqref{eq.dadRi2} with $\Im\{C_0\}$ evaluated locally at each $(Re,F)$ is required. Using this local evaluation, the first-order buoyancy correction agrees closely with the solution of the buoyant eigenvalue problem \eqref{eq:residual_wbuo}, with deviations in the neutral curve within $\sim\!3\%$. Panel (b) shows isolines of constant phase speed $c_\mathit{r}=0.32$ in the $Re$--$F$ plane. The phase speed computed using $\Re\{C_0\}$ agrees closely with the solution from the buoyant eigenvalue problem. According to the buoyancy-induced phase-speed modification, $\delta c_\mathit{r}=-\omega\Re\{C_0\}Ri/\Re\{\alpha_0\}^2$, we obtain $\delta c_\mathit{r}>0$ since $\Re\{C_0\}>0$ in this case, indicating that unstable stratification increases the phase-speed of the unstable mode (Mode II). These results confirm that the first-order perturbation theory remains valid even when large density variations ($\Delta\rho^*/\rho^*_\infty \approx \mathcal{O}(1)$) are present under strong non-ideal gas effects, amplifying buoyancy effects near the pseudo-critical point. The key requirement is the local evaluation of $C_0$ to capture the sharp thermophysical property gradients.
\begin{figure}
\centering
\includegraphics[angle=-0,trim=0 0 0 0, clip,width=0.88\textwidth]{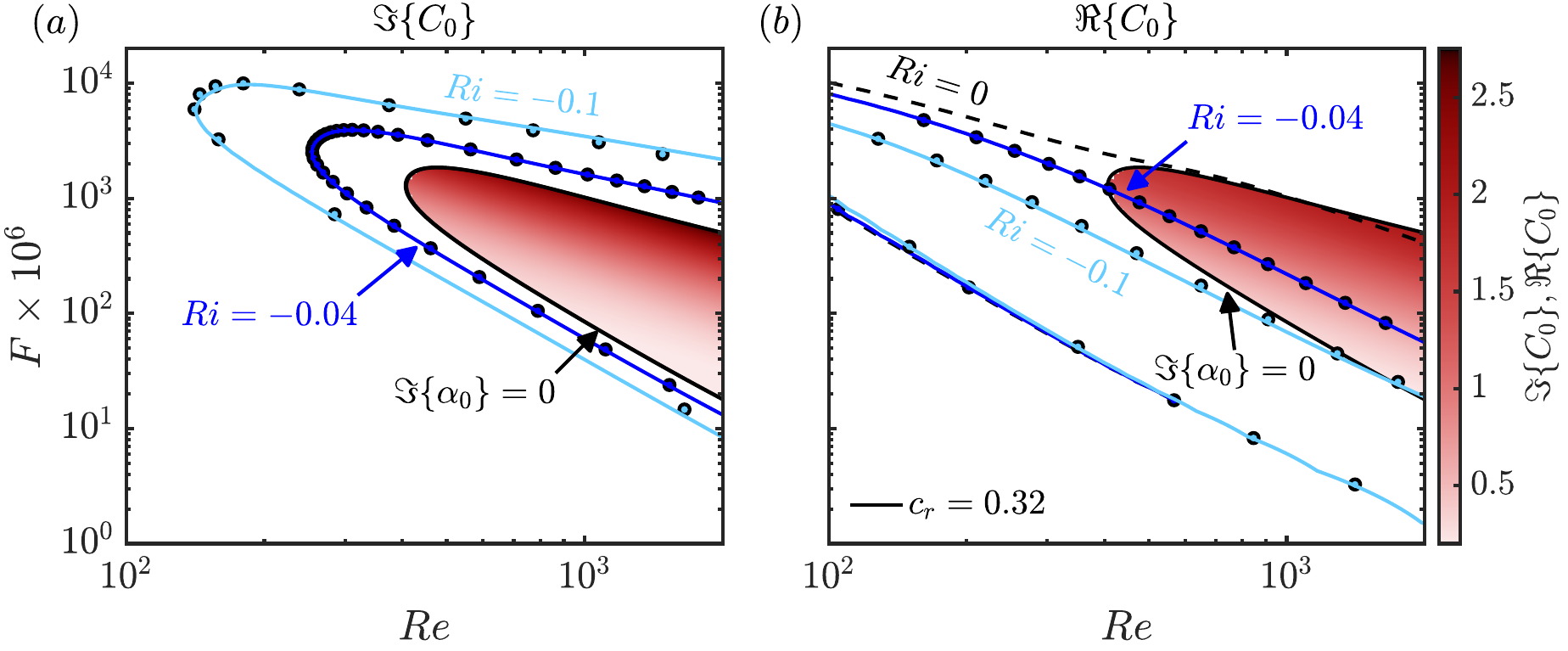}
\captionsetup{justification=justified}
  \caption{Wall-heating case at two levels of unstable stratification ($Ri=-0.037$ at $M_\infty=0.05$ and $Ri=-0.1$ at $M_\infty=0.03$). (a) Neutral-stability curves in the $Re$--$F$ plane. (b) Isolines of constant phase speed $c_\mathit{r}=0.32$ in the $Re$--$F$ plane. Symbols (\textcolor{black}{$\circ$}) show results from the buoyant eigenvalue problem \eqref{eq:residual_wbuo}; solid lines denote the first-order correction \eqref{eq.dadRi2}. The black dashed line in (b) shows the neutrally buoyant case ($Ri=0$). Background contours show (a) $\Im\{C_0\}$ and (b) $\Re\{C_0\}$ at $M_\infty=0.05$ (contours at $M_\infty=0.03$ are nearly identical). }
\label{fig9}
\end{figure}

\begin{figure}
\centering
\includegraphics[angle=-0,trim=0 0 0 0, clip,width=0.97\textwidth]{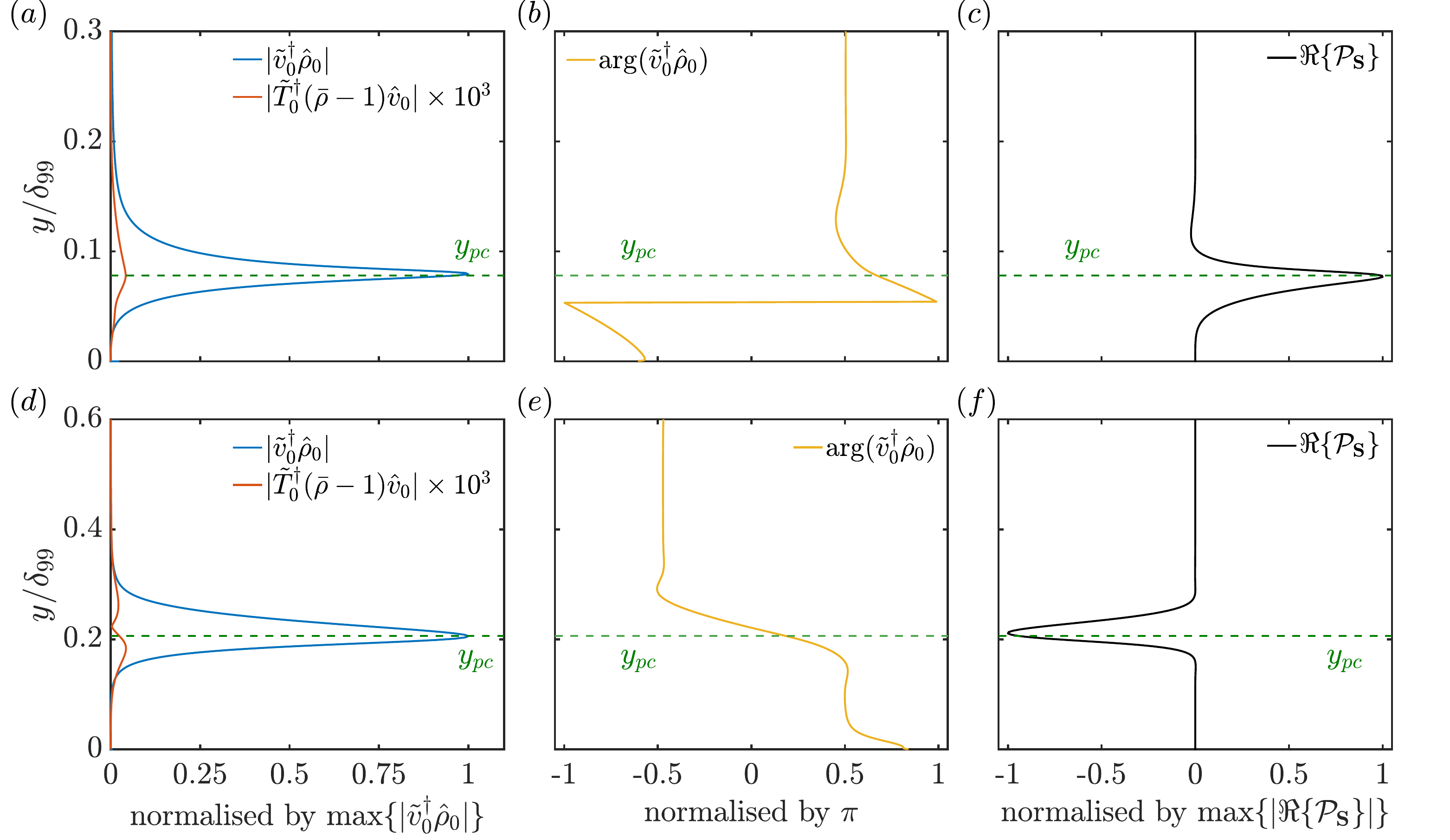}
\captionsetup{justification=justified}
\caption{Buoyancy sensitivity and production for (a--c) wall heating and (d--f) wall cooling. (a,d) OB correction $|\tilde{v}_0^{\dagger}\hat{\rho}_0|$ (blue) 
and NOB correction $|T_0^{\dagger}(\bar{\rho}-1)\tilde{v}_0|\times10^3$ (red) to $C_0$ 
from \eqref{eq:phys_C02}, both normalised by 
$\max\{|\tilde{v}_0^{\dagger}\hat{\rho}_0|\}$. (b,e) Phase difference $\arg(\tilde{v}_0^{\dagger}\hat{\rho}_0)/\pi$. (c,f) Normalised local buoyancy 
production $\Re\{\mathcal{P}_\mathbf{S}\}$. The horizontal green dashed line indicates the pseudo-critical point $y_\mathit{pc}$. In both cases, the eigenfunctions are extracted at $Re=1000$ and $F=100\times10^{-6}$.}
\label{fig10}
\end{figure}

In contrast to the weakly stratified base flows in \S\,\ref{sec:4}, where $C_0$ is dominated by the OB term in \eqref{eq:phys_C0}, the strongly stratified supercritical cases in figure~\ref{fig8} may introduce a NOB correction to the buoyancy sensitivity $C_0$. To assess its relevance, figures~\ref{fig10}(a,d) compare the wall-normal distributions of the two integrand terms in \eqref{eq:phys_C02} for the wall heating (panel a) and wall cooling (panel d). In both cases, the analysis is performed at $Re=1000$ and $F=100 \times 10^{-6}$, corresponding to the unstable mode in each configuration (see figures~\ref{fig9} and~\ref{figB}). The NOB term, $\tilde{T}_0^{\dagger}(\bar{\rho}-1)\hat{v}_0$, is confined near $y_\mathit{pc}$ and is negligible compared with the OB correction, $\tilde{v}^{\dagger}_0\hat{\rho}_0$. The dominance of the OB term occurs because $\hat{\rho}_0$ peaks sharply around $y_\mathit{pc}$, where $\hat{T}_0$ remains at least an order of magnitude smaller, as shown in \citet{Ren2019b,Boldini2025b}. Consequently, the buoyancy sensitivity $C_0$ can be accurately approximated by $C_{0,\mathrm{OB}}$ alone in both cases (figures~\ref{fig9} and~\ref{figB}). Note that this analysis is based on the eigenfunctions of the neutrally buoyant boundary layers (subscript $0$), as required by the first-order perturbation framework. Direct comparisons (not shown) confirm that the corresponding buoyant eigenfunctions do not differ significantly in shape.
In addition, the relation between density perturbation $\hat{\rho}_0$ and adjoint wall-normal velocity perturbation $\tilde{v}^{\dagger}_0$ determines both the magnitude and sign of the first-order buoyancy-production correction $\mathcal{P}^{(1)}_\mathbf{S}$ \eqref{eq:KE_BProd_FOC2} in the kinetic-energy budget \eqref{eq:KE}. Using the buoyancy-production term \eqref{eq:Prod_C0_Ri} and the OB term to $C_0$ \eqref{eq:phys_C02}, we write the real part of $\mathcal{P}_\mathbf{S}$ (net spatial growth) as
\begin{equation}
    \Re\{\mathcal{P}_\mathbf{S}\}=-Ri \, \Re\{C_{0,\mathrm{OB}}\}=-Ri\dfrac{\rho^*_\infty}{\Delta\rho^*}\int_0^\infty|\tilde{v}^{\dagger}_0\hat{\rho}_0|\cos(\phi)\,\mathrm{d}y,
    \label{eq:COB_final}
\end{equation}
where $\phi=\arg(\tilde{v}^{\dagger}_0 \hat{\rho}_0)$ is the phase difference between $\tilde{v}^{\dagger}_0$ and $\hat{\rho}_0$ shown in figures~\ref{fig10}(b,e) for wall heating and cooling. The phase difference, normalised by $\pi$, is close to $1$ at $y_\mathit{pc}$ for wall heating (panel b), while it approaches $0$ at $y_\mathit{pc}$ for wall cooling (panel e). Consequently, at the pseudo-critical point, when also considering $|\tilde{v}^{\dagger}_0\hat{\rho}_0|$ shown in figures~\ref{fig10}(a,d), wall heating (nearly out of phase, $\cos(\phi)\approx -1$) yields a positive buoyancy production ($\Re\{\mathcal{P}_\mathbf{S}\}>0$), whereas wall cooling (nearly in phase, $\cos(\phi)\approx 1$) yields a negative buoyancy production ($\Re\{\mathcal{P}_\mathbf{S}\}<0$). We emphasise that this behaviour concerns only the perturbation buoyancy-production term; the net spatial amplification trends result from the complete kinetic-energy budget (see Appendix~\ref{sec:appKE}). Remarkably, the nature of base flow and residual operator $\mathcal{R}$ \eqref{eq:residual_wbuo} is strongly NOB due to large density variations, and the underlying instability is driven by a mixed shear--baroclinic mechanism studied without gravitational effects \citep{Bugeat2024,Boldini2025b}, generating two out-of-phase waves around $y_\mathit{pc}$. The buoyancy-induced modification of this instability (quantified by $C_0$) acts as a localised modulation around the pseudo-critical point (not shown) and can be accurately captured using only the OB contribution (see \eqref{eq:phys_C0}). 

\begin{figure}
\centering
\includegraphics[angle=-0,trim=0 0 0 0, clip,width=0.85\textwidth]{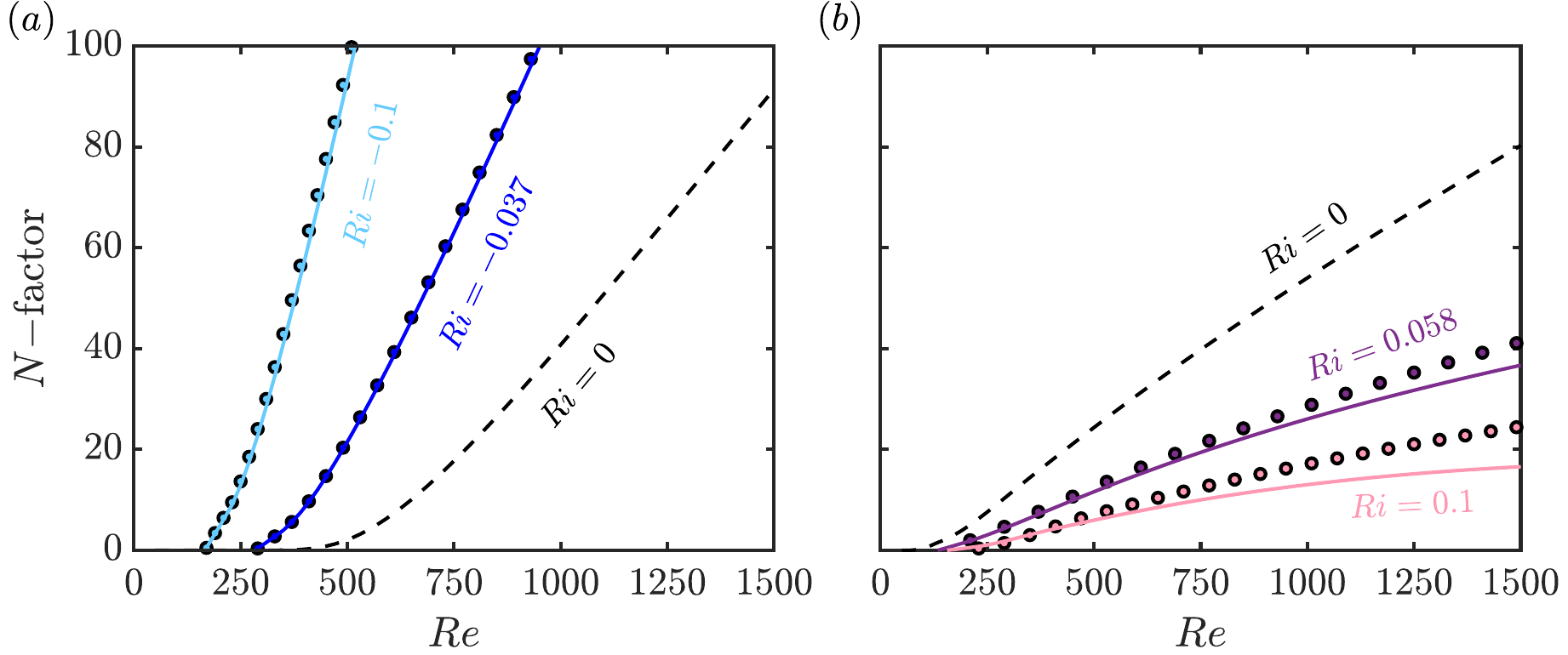}
\captionsetup{justification=justified}
  \caption{$N$-factor envelopes for (a) wall-heating cases at $Ri=-0.037$ ($M_\infty=0.05$) and $Ri=-0.1$ ($M_\infty=0.03$), and (b) wall-cooling cases at $Ri=0.058$ ($M_\infty=0.2$) and $Ri=0.1$ ($M_\infty=0.15$). Symbols (\textcolor{black}{$\circ$}) show results from the buoyant eigenvalue problem \eqref{eq:residual_wbuo}; solid lines denote the first-order correction \eqref{eq.dadRi2}. The black dashed lines indicate the neutrally buoyant reference cases ($Ri=0$).}
\label{fig11}
\end{figure}
Figure~\ref{fig11} compares the $N$-factor envelopes obtained from the buoyant eigenvalue problem \eqref{eq:residual_wbuo} with those from the first-order buoyancy correction \eqref{eq.dadRi2}. Panel (a) shows excellent agreement for the wall-heating cases at $Ri=-0.037$ ($M_\infty=0.05$) and $Ri=-0.1$ ($M_\infty=0.03$), where unstable stratification promotes earlier transition onset and larger $N$-factors compared to the neutrally buoyant reference case. Panel (b) presents the wall-cooling cases at $Ri=0.058$ ($M_\infty=0.2$) and $Ri=0.1$ ($M_\infty=0.15$). The latter case uses the same thermodynamic conditions as the case in figure~\ref{fig8}(b) but at a lower Mach number resulting in a larger stratification. In contrast to wall heating, stable stratification in wall cooling not only delays transition but also significantly reduces the integral amplification. Some deviations in $N$-factor predictions appear at large $Re$, especially for $Ri=0.1$, where higher-order buoyancy effects become non-negligible beyond the first-order buoyancy correction.

\section{Conclusions}
\label{sec:6}

We develop a perturbative, adjoint-based residual framework to predict buoyancy effects on modal instability in stratified boundary layers, within the general non-Oberbeck--Boussinesq (NOB) formulation. The first-order correction, $\delta \alpha=C_0Ri$, accurately predicts buoyancy-induced shifts of neutral curves, growth rates, phase speeds, and eigenfunctions, for both stable and unstable stratifications at moderate Richardson number. Importantly, the buoyancy sensitivity $C_0$ not only captures these shifts from the neutrally buoyant reference case but also governs the buoyancy-production term in the perturbation kinetic-energy budget, thereby linking buoyancy-induced energy transfer directly to modal growth.

For Tollmien--Schlichting waves in horizontal weakly stratified flows, the buoyancy sensitivity $C_0$ varies only weakly across the neutral stability curve, and an averaged $\bar{C}_0$ provides equally reliable $N$-factor envelopes. For air 
($Pr \approx 0.7$), $\Im\{\bar{C}_0\} \approx 0.1$, while for water 
($Pr \approx 7$), $\Im\{\bar{C}_0\} \approx 0.18$. Parametric sweeps show that $\bar{C}_0$ (i) depends strongly on Prandtl number, peaking near water-like values and changing sign at higher $Pr$, indicating that unstable stratification ($Ri<0$) can have a stabilising effect on the hydrodynamic instability; and (ii) is relatively insensitive to variations in wall-to-free-stream temperature ratio and Mach number in the low-speed ($M\lesssim0.1$) ideal-gas regime. Despite the strong Prandtl number dependence, the first-order buoyancy correction performs significantly well in predicting the $N$-factor across various temperature ratios and low Mach numbers.

Under non-ideal gas conditions at supercritical pressure, the first-order correction captures buoyancy-induced modifications of linear instability. Due to the strong base-flow stratification under pseudo-boiling conditions, with $\Delta \rho^*/\rho^*_\infty \approx \mathcal{O}(1)$, an averaged $\bar{C}_0$ is no longer accurate and instead a local $C_0$ must be used to account for sharp thermophysical property gradients near the pseudo-critical point. The sensitivity $C_0$ remains dominated by its Oberbeck--Boussinesq contribution, $\tilde{v}^{\dagger}_0\hat{\rho}_0$, which peaks at the pseudo-critical point due to strong density fluctuations. Remarkably, despite the strongly NOB underlying base flow and stability operator, the buoyancy modification of the instability is controlled by only the Oberbeck--Boussinesq contribution. At the pseudo-critical point, the phase relationship between density and wall-normal velocity perturbations determines the sign of buoyancy production: wall heating (perturbations nearly out of phase) produces positive buoyancy production (destabilising), while wall cooling (perturbations nearly in phase) produces negative buoyancy production (stabilising), significantly modifying the underlying inflectional inviscid instability (Mode II) that has been thus far studied without gravitational effects. While their quantitative impact depends on the magnitude of the Richardson number, these effects highlight the need to carefully assess buoyancy contributions in transitional flows at supercritical pressure under pseudo-boiling conditions.

In summary, we have developed a simple yet effective framework to predict buoyancy effects on modal stability in variable-property flows. The first-order correction directly determines the buoyancy-production term in the perturbation kinetic-energy budget and accurately captures how stratification modifies instability characteristics across a wide parameter space, from weakly stratified ideal-gas flows to strongly stratified non-ideal flows at supercritical pressure. The approach requires only a single adjoint calculation, enabling efficient parametric studies that would otherwise require repeated stability analyses at each stratification level. Our model is valid for moderate buoyancy effects, $|Ri| \ll 1$, and provides accurate modal stability corrections even in strongly stratified flows with $\mathcal{O}(1)$ density variations. It thus provides a foundation for further analysis of flow instabilities in stratified boundary layers and can be applied to other flow configurations. Future studies will focus on investigating buoyancy effects in the nonlinear regime, with the present linear framework providing guidance for targeted direct numerical simulations of horizontal stratified boundary layers.

\backsection[Acknowledgements]{The authors acknowledge the use of computational resources of the DelftBlue supercomputer, provided by the Delft High Performance Computing Centre (\url{https://www.tudelft.nl/dhpc}).}

\backsection[Funding]{This work was funded by the European Research Council grant no.~ERC-2019-CoG-864660, Critical.}

\backsection[Declaration of interests]{The authors report no conflict of interest.}

\backsection[Data availability statement]{The data that support the findings of this study are available on demand.}

\backsection[Author ORCIDs]{\\
P.~C.~Boldini, \url{https://orcid.org/0000-0003-0868-5895}; \\
R.~Hirai, \url{https://orcid.org/0000-0001-6644-9693}; \\
B.~Bugeat, \url{https://orcid.org/0000-0001-5420-7531}; \\
R.~Pecnik, \url{https://orcid.org/0000-0001-6352-6323}}

\appendix

\section{Stability matrices for linear stability analysis of buoyant flows} \label{sec:appA}
The base-flow matrices $\mathcal{L}_t$, $\mathcal{L}_x$, $\mathcal{L}_y$, $\mathcal{L}_q$, $\mathcal{V}_{xx}$, $\mathcal{V}_{yy}$, and $\mathcal{V}_{xy}$ (each $4\times4$), for the linearised stability equations in \eqref{eq:LinearNS}, are listed below:
\begin{align}
  \left.
    \begin{array}{rl}
&\mathcal L_{t}(1,1)=1, \\[2ex]
&\mathcal L_{t}(2,2)=\mathcal L_{t}(3,3)=\bar{\rho}  \\[2ex]
&\mathcal L_{t}(4,1)= \bar{\rho} \bar{e}_{\bar{\rho}}, \quad \mathcal L_{t}(4,4)= \bar{\rho} \bar{e}_{\bar{T}}, 
        \end{array}
  \right\}
\end{align}
\begin{align}
  \left.
    \begin{array}{rl}
&\mathcal L_{x}(1,1)=\bar{u}, \quad \mathcal L_{x}(1,2)=\bar{\rho}, \\[1.5ex]
&\mathcal L_{x}(2,1)=\bar{p}_{\bar{\rho}}, \quad \mathcal L_{x}(2,2)=\bar{\rho} \bar{u},  \quad \mathcal L_{x}(2,3)=-  \dfrac{1}{Re}\bar{\mu}_y, \quad \mathcal L_{x}(2,4)=\bar{p}_{\bar{T}}, \\[1.5ex]
&\mathcal L_{x}(3,1)= -  \dfrac{1}{Re} \bar{\mu}_{\bar{\rho}} \bar{u}_y, \quad \mathcal L_{x}(3,2)= - \dfrac{1}{Re} \bar{\lambda}_y, \\[2.0ex]
&\mathcal L_{x}(3,3)= \bar{\rho} \bar{u}, \quad \mathcal L_{x}(3,4)= -  \dfrac{1}{Re} \bar{\mu}_{\bar{T}} \bar{u}_y, \\[2.0ex]
&\mathcal L_{x}(4,1)= \bar{\rho}\bar{u} \bar{e}_{\bar{\rho}}, \quad \mathcal L_{x}(4,2)= \bar{p}, \quad \mathcal L_{x}(4,3)= - \dfrac{2 \bar{\mu}}{Re} \bar{u}_{y}, \quad \mathcal L_{x}(4,4)= \bar{\rho}\bar{u} \bar{e}_{\bar{T}}, 
  \end{array}
  \right\}
\end{align}

\begin{align}
  \left.
    \begin{array}{rl}
&\mathcal L_{y}(1,3)=\bar{\rho}, \\[1.5ex]
&\mathcal L_{y}(2,1)=- \dfrac{1}{Re} \bar{\mu}_{\bar{\rho}} \bar{u}_y,  \quad \mathcal L_{y}(2,2)=- \dfrac{1}{Re} \bar{\mu}_y,  \quad \mathcal L_{y}(2,4)=- \dfrac{1}{Re} \bar{\mu}_{\bar{T}} \bar{u}_y,  \\[2ex]
&\mathcal L_{y}(3,1)=\bar{p}_{\bar{\rho}}, \quad \mathcal L_{y}(3,3)=- \dfrac{1}{Re} \bar{\lambda}_y - \dfrac{2}{Re} \bar{\mu}_y, \quad \mathcal L_{y}(3,4)=\bar{p}_{\bar{T}}, \\[2ex]
&\mathcal L_{y}(4,1)= - \dfrac{1}{Re Ec_\infty Pr_\infty} \bar{T}_y \bar{\kappa}_{\bar{\rho}} , \quad \mathcal L_{y}(4,2)= - \dfrac{2\bar{\mu}}{Re} \bar{u}_y \\[2ex]
&\mathcal L_{y}(4,3)=\bar{p},  \quad \mathcal L_{y}(4,4)= - \dfrac{1}{Re Ec_\infty Pr_\infty} \left( \bar{T}_y \bar{\kappa}_{\bar{T}} +  \bar{\kappa}_y \right), 
      \end{array}
  \right\}
\end{align}

\begin{align}
  \left.
    \begin{array}{rl}
&\mathcal{L}_{q}(1,3)= \bar{\rho}_{y}, \\[1.5ex]
&\mathcal L_{q}(2,1)=-  \dfrac{1}{Re} \bar{\mu}_{\bar{\rho}}  \bar{u}_{yy} -  \dfrac{1}{Re} \bar{u}_y \left[ \bar{\mu}_{\bar{\rho}\bar{\rho}} \bar{\rho}_y + \bar{\mu}_{\bar{\rho}\bar{T}} \bar{T}_y \right]  , \\[2ex]
&\mathcal{L}_{q}(2,3)=\bar{\rho} \bar{u}_y ,  \\[1.5ex] 
&\mathcal L_{q}(2,4)=-  \dfrac{1}{Re} \bar{\mu}_{\bar{T}}  \bar{u}_{yy} -  \dfrac{1}{Re} \bar{u}_y \left[ \bar{\mu}_{\bar{T}\bar{T}} \bar{T}_y + \bar{\mu}_{\bar{\rho}\bar{T}} \bar{\rho}_y  \right] , \\[2ex]
&\mathcal L_{q}(3,1)=\bar{p}_{\bar{\rho}\bar{\rho}}\bar{\rho}_y+\bar{p}_{\bar{\rho}\bar{T}}\bar{T}_y, \quad \mathcal L_{q}(3,4)=\bar{p}_{\bar{T}\bar{T}}\bar{T}_y+\bar{p}_{\bar{\rho}\bar{T}}\bar{\rho}_y, \\[2ex]
&\mathcal L_{q}(4,1)=-  \dfrac{1}{Re} \bar{\mu}_{\bar{\rho}}   \bar{u}^2_y-  \dfrac{1}{Re Ec_\infty Pr_\infty} \left[ \bar{T}_{yy}  \bar{\kappa}_{\bar{\rho}} + \bar{T}_y  \left[ \bar{\kappa}_{\bar{\rho}\bar{T}} \bar{T}_y + \bar{\kappa}_{\bar{\rho}\bar{\rho}} \bar{\rho}_y \right]  \right], \\[2ex]
&\mathcal L_{q}(4,3)=\bar{\rho} \bar{e}_y, \\[1.5ex]
&\mathcal L_{q}(4,4)=-  \dfrac{1}{Re} \bar{\mu}_{\bar{T}}   \bar{u}^2_y-  \dfrac{1}{Re Ec_\infty Pr_\infty} \left[ \bar{T}_{yy}  \bar{\kappa}_{\bar{T}} +  \bar{T}_y \left[ \bar{\kappa}_{\bar{\rho}\bar{T}} \bar{\rho}_y + \bar{\kappa}_{\bar{T}\bar{T}} \bar{T}_y \right]   \right],   
              \end{array}
  \right\}
\end{align}

\begin{align}
  \left.
    \begin{array}{rl}
 &\mathcal V_{xx}(2,2)= \mathcal V_{yy}(3,3)= -\dfrac{ \bar{\lambda}}{Re}  -\dfrac{ 2\bar{\mu}}{Re}, \\[2ex]
 &\mathcal V_{xx}(3,3)= -\dfrac{ \bar{\mu}}{Re}, \quad \mathcal V_{yy}(2,2)= -\dfrac{ \bar{\mu}}{Re}, \\[2ex]
&\mathcal V_{xx}(4,4)=\mathcal V_{yy}(4,4)= -\dfrac{ \bar{\kappa}}{Re Ec_\infty Pr_\infty},  \\ [2ex]
&\mathcal V_{xy}(2,3)=\mathcal V_{xy}(3,2)=-\dfrac{ \bar{\lambda}}{Re}  -\dfrac{ \bar{\mu}}{Re}. 
              \end{array}
  \right\}
\end{align}
For simplicity, the derivative of a thermodynamic quantity with respect to $\bar{T}$ at constant $\bar{\rho}$ is denoted as $(\cdot)_{\bar{T}}$ and $(\cdot)_{\bar{T}\bar{T}}$ instead of $\partial/\partial \bar{T}|_{\bar{\rho}}$ and $\partial^2/\partial \bar{T}^2|_{\bar{\rho}\bar{\rho}}$; conversely, derivatives with respect to $\bar{\rho}$ at constant $\bar{T}$ are written as $(\cdot)_{\bar{\rho}}$ and $(\cdot)_{\bar{\rho}\bar{\rho}}$ instead of $\partial/\partial \bar{\rho}|_{\bar{T}}$ and $\partial^2/\partial \bar{\rho}^2|_{\bar{T}\bar{T}}$. With the same notation, the first- and second-order derivatives in the wall-normal direction $\mathrm{d}(\cdot)/\mathrm{d}y$ and $\mathrm{d}^2(\cdot)/\mathrm{d}y^2$ are expressed as $(\cdot)_\mathit{y}$ and $(\cdot)_\mathit{yy}$, respectively. Note that for an ideal gas: $\bar{\mu}_{\bar{\rho}}=\bar{\mu}_{\bar{\rho}\bar{\rho}}=\bar{\mu}_{\bar{\rho}\bar{T}}=\bar{\kappa}_{\bar{\rho}}=\bar{\kappa}_{\bar{\rho}\bar{\rho}}=\bar{\kappa}_{\bar{\rho}\bar{T}}=\bar{e}_{\bar{\rho}}=\bar{p}_{\bar{\rho}\bar{\rho}}=0$. The non-zero elements of the operator perturbation $\delta \mathcal{A}$ in \eqref{eq:C_term} are located in $\mathcal{L}_{q,\mathbf{S}}$ as follows:
\begin{align}
  \left.
    \begin{array}{rl}
&\mathcal L_{q,\mathbf{S}}(3,1)=-Ri\dfrac{\rho^*_\infty}{\Delta\rho^*}, \quad \mathcal L_{q,\mathbf{S}}(4,3)=-Ri\dfrac{\rho^*_\infty}{\Delta\rho^*}(\bar{\rho}-1). 
              \end{array}
  \right\}
\end{align}

\section{First-order buoyancy correction of the disturbance eigenfunctions} \label{sec:appB}

\begin{figure}
\centering
\includegraphics[angle=-0,trim=0 0 0 0, clip,width=0.85\textwidth]{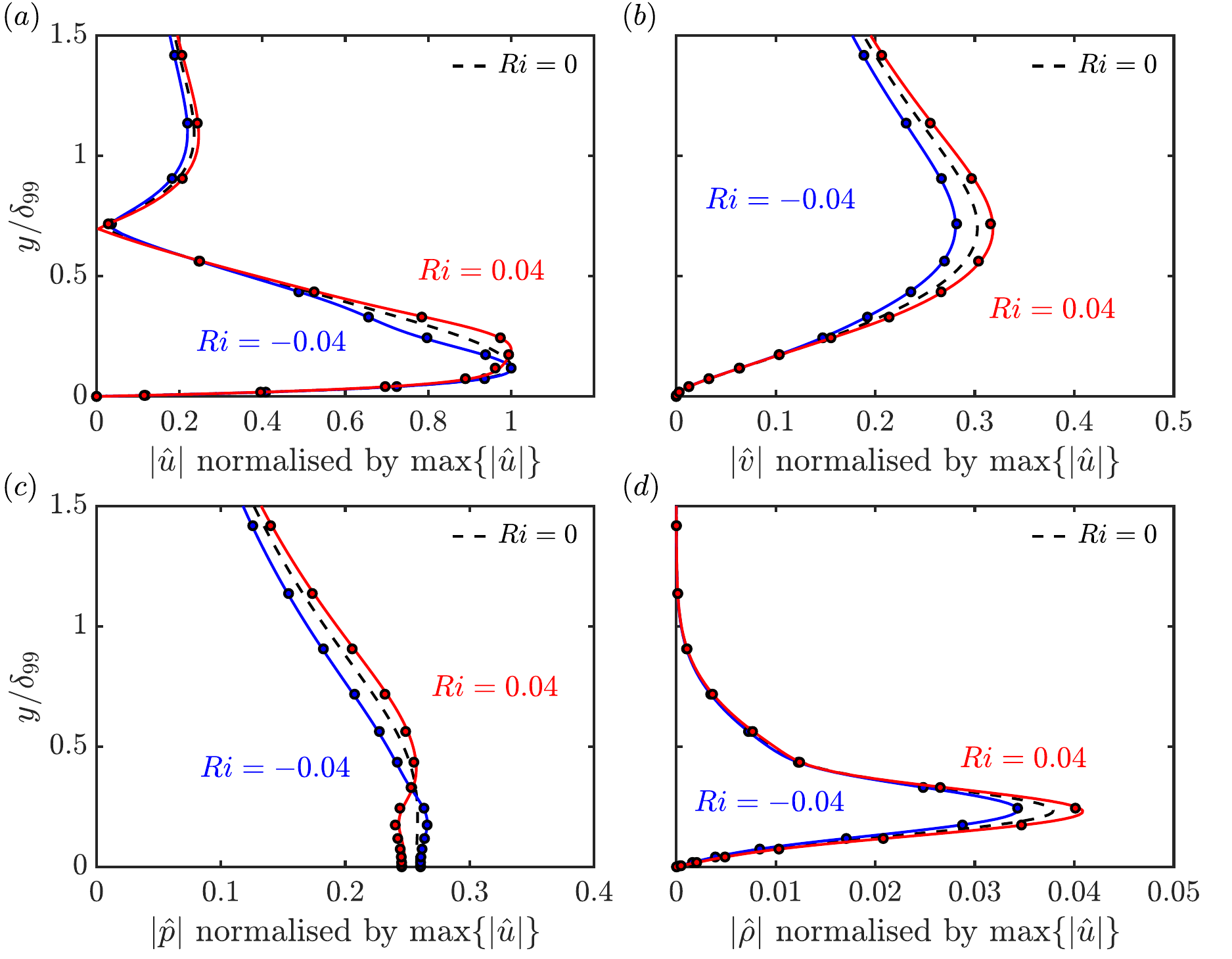}
\captionsetup{justification=justified}
  \caption{Wall-normal eigenfunctions of (a) streamwise velocity $u$, (b) wall-normal velocity $v$, (c) pressure $p$, and (d) density $\rho$ normalised by the respective $\max\{|\hat{u}|\}$, for $Ri=[-0.04,0.04]$, $Re=1000$, and $F=45 \times 10^{-6}$ (black pentagram in figure~\ref{fig3}). Symbols (\textcolor{black}{$\circ$}) show results from the buoyant eigenvalue problem \eqref{eq:residual_wbuo}; solid lines denote the first-order correction \eqref{eq.dadRi2}. The neutrally buoyant case at $Ri=0$ is indicated with a black dashed line. }
\label{figA}
\end{figure}

Figure~\ref{figA} compares the eigenfunctions of streamwise velocity $u$ (panel a), wall-normal velocity $v$ (panel b), pressure $p$ (panel c), and density $\rho$ (panel d), normalised by $\max{|\hat{u}|}$, obtained from the first-order buoyancy correction with those from the buoyant eigenvalue problem \eqref{eq:residual_wbuo}. Excellent agreement is achieved. Both stable ($Ri=0.04$) and unstable ($Ri=-0.04$) stratification are considered at $Re=1000$ and $F=45\times10^{-6}$ (black pentagram in figure~\ref{fig3}), with the neutrally buoyant case shown for reference.
Stable stratification ($Ri>0$) produces an upward shift of the $\max{|\hat{u}|}$ location in agreement with \citet{Thummar2024}, whereas for unstable stratification ($Ri<0$) the peak moves closer to the wall. The larger relative amplitude of $\hat{v}$ under stable stratification reflects the enhanced vertical motion needed to overcome the stabilising buoyancy force, whereas destabilising buoyancy assists vertical displacement, requiring a smaller peak in $\hat{v}$. The pressure eigenfunction $\hat{p}$ shifts its peak from the wall to the central region of the boundary layer, where a larger pressure response is required to balance the stabilising buoyancy force, accompanied by an increase in $\hat{\rho}$.

\section{Wall cooling under pseudo-boiling conditions}
\label{sec:appD}

\begin{figure}
\centering
\includegraphics[angle=-0,trim=0 0 0 0, clip,width=0.88\textwidth]{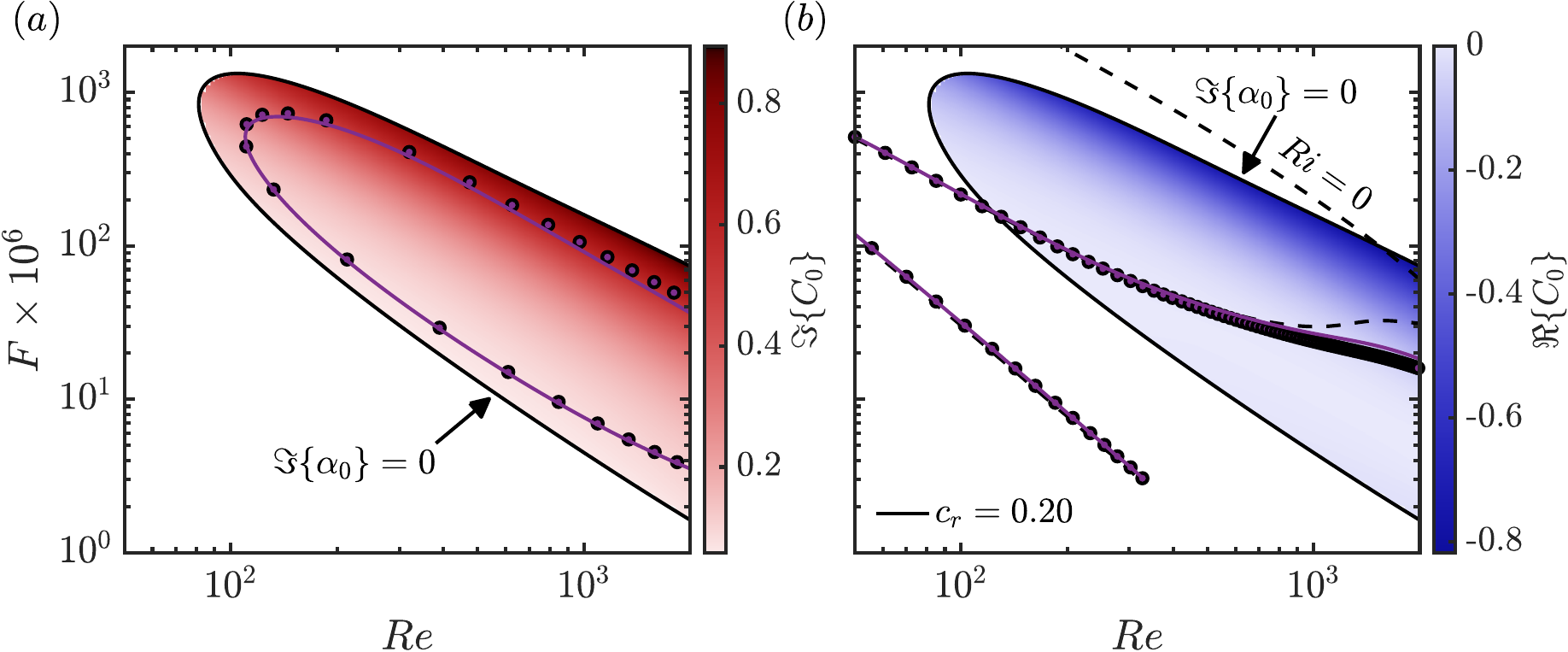}
\captionsetup{justification=justified}
    \caption{Wall-cooling case at $Ri=0.058$. (a) Neutral-stability curves in the $Re$--$F$ plane. (b) Isolines of constant phase speed $c_\mathit{r}=0.20$ in the $Re$--$F$ plane. Symbols (\textcolor{black}{$\circ$}) show results from the buoyant eigenvalue problem \eqref{eq:residual_wbuo}; solid lines denote the first-order correction \eqref{eq.dadRi2}. the black dashed line in (b) shows the neutrally buoyant case ($Ri=0$). Background contours show (a) $\Im\{C_0\}$ and (b) $\Re\{C_0\}$.}
\label{figB}
\end{figure}

We examine the linear stability of the wall-cooling case ($Ri=0.058$) at supercritical pressure under pseudo-boiling conditions in figure~\ref{figB}, whose base flow is shown in figure~\ref{fig8}(b). This flow configuration also generates an inflectional inviscid modal instability similar to Mode-II instability. Panel (a) shows the neutral-stability curve of the neutrally buoyant and stably stratified boundary layers in the $Re$--$F$ plane. With $\Im\{C_0\}>0$, the stable stratification shifts the neutral curve toward higher $Re$, delaying instability. As in the wall-heating case (figure~\ref{fig9}), an arithmetic mean of $C_0$ is not representative due to the sharp pseudo-boiling-induced gradients; the first-order buoyancy correction \eqref{eq.dadRi2} using $C_0=C_0(Re,F)$ accurately captures buoyancy effects for both neutral stability (panel a) and phase speed (panel b). Regarding the phase speed, buoyancy has minimal effect near branch I where $\Re\{C_0\}$ is small, but increases $c_\mathit{r}$ around branch II under the given stable stratification.

\section{Perturbation kinetic-energy budget in buoyant boundary layers}
\label{sec:appKE}
To demonstrate the physical link between $C_0$ and buoyancy production discussed in \S\,\ref{sec:3}, we derive the kinetic-energy budget for a buoyant boundary layer with two-dimensional perturbations in normal-mode form.
From the linearised stability equation \eqref{eq:LinearNS}, the spatial evolution of the perturbation kinetic energy $\Theta=\text{i}\alpha \int_0^\infty \bar{\rho} \bar{u}(\hat{u}\tilde{u}^{\dagger}+\hat{v}\tilde{v}^{\dagger})\,\mathrm{d}y$ can be written as:
\begin{equation}
\Theta= \mathcal{K} + \mathcal{P} + \mathcal{T} + \mathcal{V} + \mathcal{P}_\mathbf{S},
\label{eq:KE}
\end{equation}
where
\begin{equation}
    \mathcal{K}=\text{i}\omega \int_0^\infty \bar{\rho} (\hat{u}\tilde{u}^{\dagger}+\hat{v}\tilde{v}^{\dagger})\,\mathrm{d}y,
\end{equation}
\begin{equation}
    \mathcal{P} = - \int_0^{\infty} \bar{\rho} \, \bar{u}_y \, \hat{v} \tilde{u}^{\dagger} \, \mathrm{d}y, 
\end{equation}
\begin{flalign}
  \begin{aligned}
    \mathcal{T} = &- \int_0^{\infty} 
  \bigg[ \text{i} \alpha \hat{p} \tilde{u}^{\dagger} + \hat{p}_y \tilde{v}^{\dagger} \bigg] \, \mathrm{d}y,
  \end{aligned}
\end{flalign}
\begin{flalign}
  \begin{aligned}
 \mathcal{V} = &\frac{1}{Re} \int_0^{\infty}
\bigg\{ \tilde{u}^{\dagger} \bigg[ \text{i} \alpha \bar{\mu}_y \hat{v} + \bar{\mu}_{\bar{\rho}} \bar{u}_y \hat{\rho}_y + \bar{\mu}_y \hat{u}_y + \bar{\mu}_{\bar{T}} \bar{u}_y \hat{T}_y  \\
  &
  + \bar{\mu}_{\bar{\rho}} \bar{u}_{yy} \hat{\rho} 
  + \bar{u}_{y}
  \left(
    \bar{\mu}_{\bar{\rho}\bar{\rho}} \bar{\rho}_{y}
    + \bar{\mu}_{\bar{\rho}\bar{T}} \bar{T}_{y}
  \right) \hat{\rho}    \\
  &
  + \bar{\mu}_{\bar{T}} \bar{u}_{yy} \hat{T}
  + \bar{u}_{y}
  \left(
   \bar{\mu}_{\bar{T}\bar{T}} \bar{T}_{y}
    + \bar{\mu}_{\bar{\rho} \bar{T}} \bar{\rho}_{y}
  \right) \hat{T}
  \\
  &
  - \alpha^2 (2 \bar{\mu} + \bar{\lambda}) \, \hat{u}
  + \bar{\mu} \hat{u}_{yy}
  + \text{i} \alpha (\bar{\mu} + \bar{\lambda}) \hat{v}_{y} \bigg] \\
    & +\tilde{v}^{\dagger} \bigg[ \text{i} \alpha \bar{\mu}_{\bar{\rho}}  \bar{u}_{y} \hat{\rho}
  + \text{i} \alpha \bar{\lambda}_{y} \hat{u}
  + \text{i} \alpha \bar{\mu}_{\bar{T}} \bar{u}_{y} \hat{T}
  + \left(
      2 \bar{\mu}_{y}
      + \bar{\lambda}_{y}
    \right)
    \hat{v}_{y} \\
  &
  - \alpha^2 \bar{\mu} \, \hat{v}
  + (2 \bar{\mu} + \bar{\lambda}) \hat{v}_{yy}
  + \text{i} \alpha (\bar{\mu} + \bar{\lambda}) \hat{u}_{y}  \bigg]
\bigg\} \, \mathrm{d}y,
  \end{aligned}
\end{flalign}
\begin{equation}
    \mathcal{P}_\mathbf{S} = - Ri \dfrac{\rho^*_\infty}{\Delta \rho^*} \int_0^{\infty}  \hat{\rho} \tilde{v}^{\dagger} \, \mathrm{d}y.
    \label{eq:KE_BProd}
\end{equation}
The term $\mathcal{K}$ represents the temporal evolution and is purely imaginary. The terms $\mathcal{P}$, $\mathcal{T}$, $\mathcal{V}$, and $\mathcal{P}_\mathbf{S}$ 
represent shear production, thermodynamic contributions, viscous dissipation, and 
buoyancy production, respectively. The real part of the kinetic-energy budget \eqref{eq:KE} gives the spatial amplification rate of the perturbation energy. The notation follows Appendix~\ref{sec:appA}, and the pressure disturbance $\hat{p}$ is given by the first-order Taylor series $\bar{p}_{\bar{\rho}}\hat{\rho}+\bar{p}_{\bar{T}}\hat{T}$.

Applying the Taylor expansion \eqref{eq:Taylor_expansion} in the Richardson number about the neutrally buoyant reference state, the buoyancy–production term \eqref{eq:KE_BProd} can be expanded as
\begin{equation}
    \mathcal{P}_{\mathbf{S}} = \mathcal{P}^{(0)}_{\mathbf{S}}+ Ri \, \mathcal{P}^{(1)}_{\mathbf{S}} + \mathcal{O}(Ri^2), \quad \text{with} \; \mathcal{P}^{(0)}_{\mathbf{S}}=0,
    \label{eq:KE_BProd_FOC}
\end{equation}
where the first-order correction term is
\begin{equation}
    \mathcal{P}^{(1)}_{\mathbf{S}} = -\frac{\rho^*_\infty}{\Delta \rho^*}
    \int_0^{\infty} \hat{\rho}_0\tilde{v}_0^{\dagger}\,\mathrm{d}y,
    \label{eq:KE_BProd_FOC2}
\end{equation}
and the subscript $0$ denotes quantities evaluated for the neutrally buoyant reference state. Taking the real part of the buoyancy-production term, i.e.~the net buoyancy production, in \eqref{eq:KE_BProd_FOC} yields
\begin{equation}
    \Re\{\mathcal{P}_\mathbf{S}\} = - Ri \dfrac{\rho^*_\infty}{\Delta \rho^*}\Re\bigg\{\int_0^{\infty}\hat{\rho}_0 \tilde{v}_0^{\dagger} \, \mathrm{d}y\bigg\}= - Ri \, \Re\{C_{0,\mathrm{OB}}\},
    \label{eq:Prod_C0_Ri}
\end{equation}
which directly relates the buoyancy production to the buoyancy sensitivity $C_{0,\mathrm{OB}}$ defined in \eqref{eq:phys_C02}. The sign of the right-hand side of the buoyancy production \eqref{eq:Prod_C0_Ri}, $- Ri \, \Re\{C_{0,\mathrm{OB}}\}$, determines whether the buoyancy contribution is stabilising or destabilising. The overall spatial growth of the perturbation, however, follows from the complete kinetic-energy budget \eqref{eq:KE}.

\begin{figure}
\centering
\includegraphics[angle=-0,trim=0 0 0 0, clip,width=0.88\textwidth]{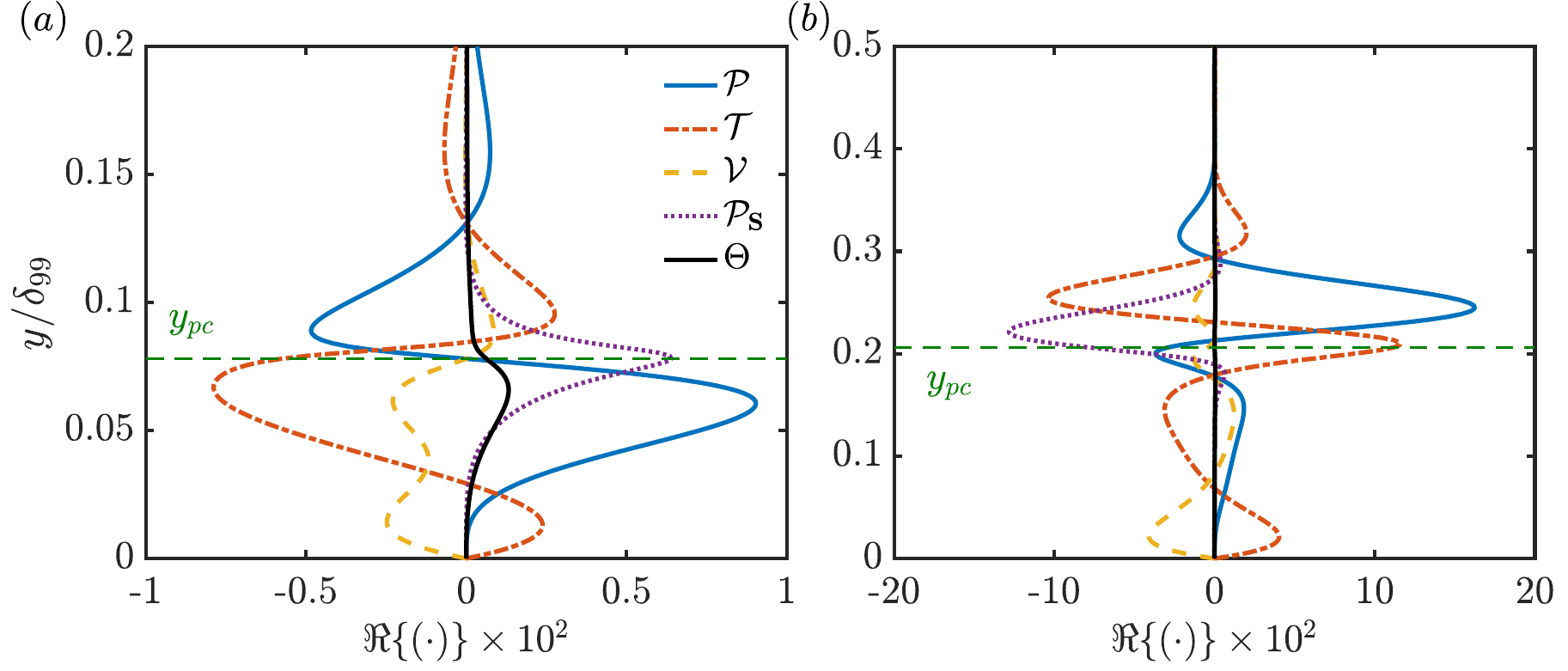}
\captionsetup{justification=justified}
  \caption{Real parts of the kinetic-energy budget terms for the boundary layers of \S\,\ref{sec:5}, under (a) wall heating and (b) wall cooling, plotted along the wall-normal direction $y/\delta_{99}$, where $\delta_{99}$ is the boundary-layer thickness. Curves show $\mathcal{P}$ (shear production), $\mathcal{T}$ (thermodynamics), $\mathcal{V}$ (dissipation), and $\mathcal{P}_\mathbf{S}$ (buoyancy production), along with the net spatial growth $\Theta$. The green dashed line indicates the pseudo-critical point $y_\mathit{pc}$. Eigenfunctions are normalised by $\max\{|\hat{u}|\}$ and extracted at $Re=1000$ and $F=100 \times 10^{-6}$.}
\label{figC}
\end{figure}

The results of the kinetic-energy budget for the boundary-layer flows at supercritical pressure in \S\,\ref{sec:5} are illustrated in figure~\ref{figC}.  The analysis is performed at $Re=1000$ and $F=100 \times 10^{-6}$ (see figures~\ref{fig9} and~\ref{figB}) for $\alpha_\mathit{i}<0$, i.e.~$\Re\{\Theta\}>0$, with panel (a) showing wall heating and panel (b) showing wall cooling. 
In the wall-heating case, $\mathcal{P}_{\mathbf{S}}$ is positive and peaks near the pseudo-critical point, as observed in figure~\ref{fig10}(c). This localised positive buoyancy production adds to the larger shear production $\mathcal{P}$, yielding positive net spatial growth $\Theta$ in the vicinity of $y_\mathit{pc}$. As a result, the Mode-II instability is further destabilised (see figure~\ref{fig9}). In contrast, for wall-cooling, $\mathcal{P}_{\mathbf{S}}$ significantly opposes the shear production around $y_\mathit{pc}$ and, overall, has a stabilising effect on the Mode-II-like instability as observed in figure~\ref{figB}. For both cases, the thermodynamic term $\mathcal{T}$ is largest at the pseudo-critical point due to its strong sensitivity to the base-flow density and temperature profiles, as well as to thermodynamic derivatives \citep{Ren2025}.

\bibliographystyle{jfm}
\bibliography{main}

@article{Zonta2018,
  author={F.~Zonta and A.~Soldati},
  title={Stably stratified wall-bounded turbulence},
  journal={Appl. Mech. Rev.},
  volume={70},
  pages={040801},
  year={2018},
}

@article{Facchini2018,
  author={G.~Facchini and B.~Favier and P.~{Le Gal} and M.~Wang and M.~{Le Bars}},
  title={The linear instability of the stratified plane {C}ouette flow},
  journal={J.~Fluid Mech.},
  volume={853},
  pages={205–234},
  year={2018},
}

@article{Gage1968,
  author={K.~S.~Gage and W.~H.~Reid},
  title={The stability of thermally stratified plane {P}oiseuille flow},
  journal={J.~Fluid Mech.},
  volume={33},
  pages={21–32},
  year={1968},
}

@article{Biau2004,
  author={D.~Biau and A.~Bottaro},
  title={The effect of stable thermal stratification on shear flow stability},
  journal={Phys.~Fluids},
  volume={16},
  issue={12},
  pages={4742–4745},
  year={2004},
}

@article{Gal2021,
  author={P.~{Le Gal} and U.~Harlander and I.~D.~Borcia and S.~{Le Dizès} and J.~Chen and B.~Favier},
  title={Instability of vertically stratified horizontal plane {P}oiseuille flow},
  journal={J.~Fluid Mech.},
  volume={907},
  pages={R1},
  year={2021},
}

@article{Sameen2007,
  author={A.~Sameen and R.~Govindarajan},
  title={The effect of wall heating on instability of channel flow},
  journal={J.~Fluid Mech.},
  volume={577},
  pages={417–442},
  year={2007},
}

@article{Parente2020,
    author={E.~Parente and J.~C.~Robinet and P.~de~Palma and S.~Cherubini},
    title={Modal and nonmodal stability of a stably stratified boundary layer flow},
    journal={Phys.~Rev.~Fluids},
    volume={5},
    pages={113901},
    year={2020}
}

@article{Hamada2023,
  author={G.~Y.~R.~Hamada and W.~R.~Wolf and D.~B.~Pitz and L.~S.~{de.~B.~Alves}},
  title={Stability and receptivity analyses of mixed convection in unstably stratified horizontal boundary layers},
  journal={J.~Fluid Mech.},
  volume={961},
  pages={A10},
  year={2023},
}

@article{Thummar2024,
  author={M.~Thummar and R.~Bhoraniya and V.~Narayanan},
  title={Stability and receptivity analyses of the heated flat-plate boundary layer with variable viscosity},
  journal={Int. J. Heat Fluid Flow},
  volume={110},
  pages={109624},
  year={2024},
}

@article{Boldini2024,
    author={P.~C.~Boldini and B.~Bugeat and J.~W.~R.~Peeters and M.~Kloker and R.~Pecnik},
    title={Transient growth in diabatic boundary layers with fluids at supercritical pressure},
    journal={Phys.~Rev.~Fluids},
    volume={9},
    pages={083901},
    year={2024}
}

@article{Caulfield2021,
  author={C.~P.~Caulfield},
  title={Layering, Instabilities, and Mixing in Turbulent Stratified Flows},
  journal={Annu. Rev. Fluid Mech.},
  volume={53},
  pages={113–145},
  year={2021},
}

@article{Boldini2025,
  title = {{CUBENS}: {A} {GPU}-accelerated high-order solver for wall-bounded flows with non-ideal fluids},
  journal = {Comput. Phys. Commun.},
  volume = {309},
  pages = {109507},
  year = {2025},
  issn = {0010-4655},
  doi = {https://doi.org/10.1016/j.cpc.2025.109507},
  author = {P.~C.~Boldini and R.~Hirai and P.~Costa and J.~W.~R.~Peeters and R.~Pecnik},
}

@ARTICLE{Malik1990,
   author       = "M.~R.~Malik",
   title        = "Numerical methods for hypersonic boundary layer stability",
   year         = "1990",
   journal      = "J.~Comput.~Phys.",
   volume       = "86",
   pages        = "376-413",
}

@book{Schlichting2003,
    author = "Schlichting, H. and Gersten, K.",
    title = "Boundary Layer Theory",
    publisher = "Springer",
    year = "2003", 
}

@book{Drazin2004,
    author = "Drazin, P.~G. and Reid, W.~H.",
    title = "Hydrodynamic Stability",
    publisher = "Cambridge University Press.",
    year = "2004", 
}

@ARTICLE{Chen2016,
   author       = "Chen, J. and Bai, Y. and Le Dizès, S.",
   title        = "Instability of a boundary layer flow on a vertical wall in a stably stratified fluid",
   year         = "2016",
   journal      = "J.~Fluid Mech.",
   volume       = "795",
   pages        = "262--277",
}

@ARTICLE{Miles1961,
   author       = "Miles, J.~W.",
   title        = "On the stability of heterogeneous shear flows",
   year         = "1961",
   journal      = "J.~Fluid Mech.",
   volume       = "10",
   issue        = "4",
   pages        = "496--508",
}

@ARTICLE{Howard1961,
   author       = "Howard, L.~N.",
   title        = "Note on a paper of {J}ohn {W}.~{M}iles",
   year         = "1961",
   journal      = "J.~Fluid Mech.",
   volume       = "10",
   issue        = "4",
   pages        = "509--512",
}

@ARTICLE{Wu1976,
   author       = "Wu, R.~S. and Cheng, K.~C.",
   title        = "Thermal instability of {B}lasius flow along horizontal plates",
   year         = "1976",
   journal      = "Intl.~J.~Heat Mass Transfer",
   volume       = "19",
   issue        = "8",
   pages        = "907--913",
}

@article{Gage1971,
  author={K.~S.~Gage},
  title={The effect of stable thermal stratification on the stability of viscous parallel flows},
  journal={J.~Fluid Mech.},
  volume={47},
  issue={1},
  pages={1--20},
  year={1971},
}

@article{Gebhart1973,
  author={B.~Gebhart},
  title={Instability, transition, and turbulence in buoyancy-induced flows},
  journal={Annu.~Rev.~Fluid Mech.},
  volume={5},
  issue={1},
  pages={213--246},
  year={1973},
}

@article{Hall1992,
  author={P.~Hall and H.~Morris},
  title={On the instability of boundary layers on heated flat plates},
  journal={J.~Fluid Mech.},
  volume={245},
  pages={367--400},
  year={1992},
}

@article{Carriere1999,
  author={P.~Carrière and P.~A.~Monkewitz},
  title={Convective versus absolute instability in mixed {R}ayleigh--{B}énard--{P}oiseuille convection},
  journal={J.~Fluid Mech.},
  volume={384},
  pages={243--262},
  year={1999},
}

@article{Hirata2015,
  author={S.~C.~Hirata and L.~S.~{de B.~Alves} and N.~Delenda and M.~N.~Ouarzazi},
  title={Convective and absolute instabilities in {R}ayleigh--{B}énard--{P}oiseuille mixed convection for viscoelastic fluids},
  journal={J.~Fluid Mech.},
  volume={765},
  pages={167--210},
  year={2015},
}

@article{Mack1984,
  author      = {L.~M.~Mack},
  title       = {Boundary-layer linear stability theory},
  journal = {AGARD  Report  No. 709: Special Course  on  Stability  and  Transition  of  Laminar  Flow.},
  note    = {{AGARD}},
  year        = {1984},
}

@article{Wang2023,
title = {Direct numerical simulation of thermal stratification of supercritical water in a horizontal channel},
journal = {Comput.~Fluids},
volume = {261},
pages = {105911},
year = {2023},
author = {W.~Wang and S.~He and C.~Moulinec and D.~R.~Emerson},
}

@article{Draskic2025,
  author={M.~Draskic and J.~Westerweel and R.~Pecnik},
  title={The stability of stratified horizontal flows of carbon dioxide at supercritical pressures},
  journal={J.~Fluid Mech.},
  volume={1012},
  pages={A17},
  year={2025},
}

@ARTICLE{Banuti2015,
   author       = "D.~T.~Banuti",
   title        = "Crossing the {W}idom-line -- {S}upercritical pseudo-boiling",
   year         = "2015",
   journal      = "J.~Supercrit.~Fluids",
   volume       = "98",
   pages        = "12–16",
}

@ARTICLE{Ren2019b,
   author       = "J.~Ren and O.~Marxen and R.~Pecnik",
   title        = {Boundary-layer stability of supercritical fluids in the vicinity of the {W}idom line},
   year         = "2019",
   journal      = "J.~Fluid Mech.",
   volume       = "871",
   pages        = "831–864",
}

@BOOK{Kato1995,
  author    = {T.~Kato},
  title     = {Perturbation Theory for Linear Operators},
  publisher = {Springer},
  year      = {1995},
}

@ARTICLE{Brunner2010,
    author = {G.~Brunner},
    title = {Applications of supercritical fluids},
    journal = {Annu.~Rev.~Chem.~Biomol.~Eng.},
    volume = {1},
    pages = {321--342},
    year = {2010}
}

@MISC{Lemmon2013,
   author       = "E.~W.~Lemmon and M.~L.~Huber and M.~O.~Mclinden",
   title        = "{NIST Standard Reference Database 23:~Reference Fluid Thermodynamic and Transport Properties - REFPROP, Version 9.1}",
   year         = "2013",
   institution  = "National Institute of Standards and Technology",
   note       = "{Available at:~\url{http://www.nist.gov/srd/nist23.cfm}}",
}

@misc{Boldini2025b,
      title={Direct numerical simulation of complete transition to turbulence with a fluid at supercritical pressure}, 
      author={P.~C.~Boldini and B.~Bugeat and J.~W.~R.~Peeters and M.~Kloker and R.~Pecnik},
      year={2025},
      eprint={2506.06703},
      archivePrefix={arXiv},
      primaryClass={physics.flu-dyn},
      url={https://arxiv.org/abs/2506.06703}, 
}

@ARTICLE{Ren2025,
   author       = "J.~Ren and Y.~Wu and X.~Mao and C.~Wang and M.~Kloker",
   title        = {Sensitivity of three-dimensional boundary layer stability to intrinsic uncertainties of fluid properties: a study on supercritical $\text{CO}_2$},
   year         = "2025",
   journal      = "J.~Fluid Mech.",
   volume       = "1007",
   pages        = "A7",
}

@article{Variale2024,
    author={D.~Variale and E.~Parente and J.~C.~Robinet and S.~Cherubini},
    title={Modal and nonmodal stability analysis of turbulent stratified channel flows},
    journal={Phys.~Rev.~Fluids},
    volume={9},
    pages={013904},
    year={2024}
}

@article{Liu2019,
    author={Y.~Liu and Y.~Wang and D.~Huang},
    title={Supercritical $\text{CO}_2$ {B}rayton cycle: a state-of-the-art review},
    journal={Energy},
    volume={189},
    pages={115900},
    year={2019}
}

@article{Guardone2024,
author = {A.~Guardone and P.~Colonna and M.~Pini and A.~Spinelli},
title = {Nonideal Compressible Fluid Dynamics of Dense Vapors and Supercritical Fluids},
journal = {Annu.~Rev.~Fluid Mech.},
volume = {56},
number = {1},
pages = {241-269},
year = {2024},
}

@ARTICLE{Wall1997,
   author       = "D. P. Wall and S. K. Wilson",
   title        = {The linear stability of flat-plate boundary-layer flow of fluid with temperature-dependent viscosity},
   year         = "1997",
   journal      = "Phys.~Fluids",
   volume       = "9",
   issue        = "10",
   pages        = "2885",
}

@article{Li2025,
title = {A review of microscale physics and macroscale convective heat transfer in supercritical fluids for energy and propulsion systems},
journal = {Appl.~Therm.~Eng.},
volume = {272},
pages = {126380},
year = {2025},
author = {Z.~Li and D.~T. Banuti and J.~Ren and J.~Lyu and H.~Wang and X.~Chu},
}

@ARTICLE{Ke2024,
   author       = "J.~Ke and S.~W.~Armfield and N.~Williamson",
   title        = {Non-{O}berbeck--{B}oussinesq effects on the linear stability of a vertical natural convection boundary layer},
   year         = "2024",
   journal      = "J.~Fluid Mech.",
   volume       = "988",
   pages        = "A44",
}

@article{Guha2018,
  title={On the inertial effects of density variation in stratified shear flows},
  author={Guha, A. and Raj, R.},
  journal={Phys.~Fluids},
  volume={30},
  number={12},
  pages={126603},
  year={2018},
}

@ARTICLE{Robinet2019,
   author       = "J.-C.~Robinet and X.~Gloerfelt",
   title        = "Instabilities in non-ideal fluids",
   year         = "2019",
   journal      = "J.~Fluid Mech.",
   volume       = "880",
   pages        = "1–4",
}

@ARTICLE{Bugeat2024,
    author="B.~Bugeat and P.~C.~Boldini and A.~M.~Hasan and R.~Pecnik",
    title="Instability in strongly stratified plane {C}ouette flow with application to supercritical fluids",
    year="2024",
    journal      = "J.~Fluid Mech.",
    volume       = "984",
    pages        = "A31",
}

\end{document}